\documentclass[aps, prd, twocolumn, showpacs, superscriptaddress, groupedaddress]{revtex4} 
\usepackage{graphicx}	
\usepackage{amssymb}
\usepackage{dcolumn}
\usepackage{color}
\usepackage{subfigure, rotating, bm, array}
\usepackage[pagebackref=false, colorlinks=true]{hyperref}
\hypersetup{
linkcolor=blue,     
citecolor=blue,     
urlcolor=blue}      
\begin{document}
\title{Timelike geodesics in Naked Singularity and Black Hole Spacetimes }
\author{Parth Bambhaniya}
\email{grcollapse@gmail.com}
\affiliation{International Center for Cosmology, Charusat University, Anand 388421, Gujarat, India}
\author{Ashok B. Joshi}
\email{gen.rel.joshi@gmail.com}
\affiliation{International Center for Cosmology, Charusat University, Anand 388421, Gujarat, India}
\author{Dipanjan Dey}
\email{dipanjandey.adm@charusat.edu.in}
\affiliation{International Center for Cosmology, Charusat University, Anand 388421, Gujarat, India}
\author{Pankaj S. Joshi}
\email{psjprovost@charusat.ac.in}
\affiliation{International Center for Cosmology, Charusat University, Anand 388421, Gujarat, India}

\date{\today}

\begin{abstract}
In this paper, we derive the solutions of orbit equations for a class of naked singularity spacetimes, and compare these with timelike orbits, that is, particle trajectories in the Schwarzschild black hole spacetime. The Schwarzschild and naked singularity spacetimes considered here can be formed as end state of a spherically symmetric gravitational collapse of a matter cloud.  We find and compare the perihelion precession of the particle orbits in the naked singularity spacetime with that of the Schwarzschild black hole. We then discuss different distinguishable physical properties of timelike orbits in the black hole and naked singularity spacetimes and implications are discussed. Several interesting differences follow from our results, including the conclusion that in naked singularity spacetimes, particle bound orbits can precess in the opposite direction of particle motion, which is not possible in Schwarzschild spacetime.


\end{abstract}


\maketitle

\section{Introduction}
The Schwarzschild spacetime is unique, static, spherically symmetric vacuum solution of Einstein equations. It represents 
a non-rotating, uncharged black hole. It is generally believed that when a star with mass greater than a critical mass (greater than about $4 M_{\odot}$) exhausts all its nuclear fuel, it then undergoes a catastrophic, continual gravitational collapse, finally terminating into a black hole. Using Einstein equations, one can show that a spherically symmetric, homogeneous, dust collapse always terminates into a Schwarzschild black hole. Oppenheimer, Snyder, and also Datt, first showed the dynamical collapse solution (OSD collapse) of a spherically symmetric, homogeneous, dust cloud \cite{OppenheimerSnyder39},\cite{Datt}. 

Further to this, more physically realistic gravitational collapse models, such as those with inhomogeneous matter distribution, or those with non-zero pressures and other scenarios have been studied in detail. It is seen, for example, 
that an inhomogeneous collapse, e.g. in a physically more realistic case with matter density higher at center, results in a strong curvature singularity which can be locally or globally visible. 
A visible or naked singularity has rather different causality structure as compared to the black hole case. 
For the homogeneous dust collapse case, the trapped surfaces and apparent horizon close to the center are formed before the formation of the central strong singularity, and therefore the singularity is always locally and globally covered. On the other hand, for inhomogeneous collapse, there are classes of initial conditions for which the central strong singularity forms before the formation of the trapped surfaces and apparent horizon. In such a case, the singularity is always at least locally naked \cite{Joshi:1993zg, Mena:1999bz, Jhingan:1996jb, Singh:1994tb}. With larger values of pressure, one can show that in asymptotic time, without forming the trapped surfaces, the collapsing matter cloud can equilibrate itself into a static, singular spacetime, which include the Joshi-Malafarina-Narayan (JMN) spacetimes, Bertrand space-times etc, which are the examples of such static, singular spacetimes formed as an end state or equilibrium state of gravitational collapse in asymptotic time \cite{Joshi:2011zm, Joshi:2013dva, Banik:2016qvf}.

There are many different catastrophic collapses which are happening in our universe at different scales. In stellar scale only baryonic matter collapse is important. On the other hand, in galactic halo scale or in galaxy cluster scale, along with baryonic matter collapse one has to consider dark matter collapse as well\cite{white, GunnGott72, Hogan:1985bc}. In these different collapsing scenarios, for various initial conditions, different types of final static spacetimes can be formed. Depending upon initial conditions, the final spacetime can be a black hole (rotating or non-rotating, charged or uncharged), JMN spacetimes, Bertrand spacetimes or other possible static spacetimes. 

Based on recent research works such as above, the current emerging scenario is, a massive star undergoing catastrophic gravitational collapse towards the end of its life cycle may terminate into a black hole or a naked singularity, depending on the initial conditions from which the collapse starts.  Both the black holes and naked singularity models would have very different physical properties and there must be some distinguishable observational signatures of those different spacetimes. Various researchers have investigated this possibility and the observational signatures of different naked singularity spacetimes have been examined in some detail. In \cite{Joshi:2013dva}, \cite{Kovacs:2010xm} the authors investigated the accretion disk properties and gravitational lensing properties of naked singularity spacetimes. In \cite{Shaikh:2018lcc}, the shadows created by a black hole and a JMN naked singularity are compared and it is shown that for some cases, the naked singularity spacetime casts a similar shadow compared to what a Schwarzschild black hole does. This is an important theoretical prediction in the context of recent observation of the shadow of M87 galactic center \cite{Akiyama:2019fyp}. In \cite{Dey:2013yga},\cite{Dey+15} astrophysical importance of Bertrand spacetimes is discussed. The gravitational lensing due to the Janis-Newman-Winicour (JNW) naked singularity is discussed in \cite{Gyulchev:2019tvk}, where authors have reported that JNW spacetime can cast a shadow which closely resembles to the shadow of a Schwarzschild black hole.  Therefore, the theoretical predictions of possible observational signatures of naked singularity spacetimes would be worth exploring.  

From such a perspective, another possibility that we explore in the present work is, comparing the particle trajectories in certain 
black hole and naked singularity spacetimes. We know that freely falling massive particles follow timelike geodesics. The curvature and causal structure of the spacetime determines the nature of these timelike geodesics or nature of the particle trajectories. Therefore, observing and studying timelike geodesics in a given geometry gives information about the causal structure of that spacetime. There are many literature where the timelike geodesics in different spacetimes are investigated \cite{Zhou, Levin:2009sk, levin1, levin, Glampedakis:2002ya, Chu:2017gvt, Dokuchaev:2015zxe, Borka:2012tj,Martinez:2019nor,Fujita:2009bp,Wang:2019rvq,Suzuki:1997by,Zhang:2018eau,Pugliese:2013zma,Farina:1993xw,Dasgupta:2012zf,Shoom}.

It may be useful to note here that recently, GRAVITY and SINFONI, which are near infrared integral field spectrometers, reported their updated data on the stellar motions near the center of our milky way \cite{M87,Eisenhauer:2005cv,center1}. These observations can give the information about the nature of the central object (SGR-A*) in our Milkyway galaxy, which is considered to be a super massive black hole with the mass of about $10^6 M_{\odot}$. In such a context, it would be important to do a comparative study of the nature of the timelike geodesics in different black hole and singularity geometries, and how they possibly differ from each other.

In this paper, we discuss the timelike geodesic in the 1st and 2nd kind of JMN (JMN-1 and JMN-2) spacetimes. JMN-1 spacetime is an anisotropic fluid solution of Einstein equations with zero radial pressure. On the other hand, JMN-2 spacetime is an isotropic fluid solution of Einstein equations. Both of these spacetimes have a central naked singularity. We then compare these trajectories with those for 
a Schwarzschild black hole.

In black hole physics, while we study different black hole geometries,
it is important to study the underlying physical processes 
that would give rise to such black holes. In astrophysical settings, it is the gravitational collapse of a massive star that would possibly create a black hole. For example, a spherical homogeneous collapse would create a Schwarzschild black hole, which has an event horizon and a singularity at the center. A physically more realistic class of collapse models, with non-zero pressures and inhomogeneities in the matter distribution included were studied in 
\cite{joshi:1993,Satin:2014ima,Joshi:2012mk}.
The final outcome of the collapse asymptotically then was obtained as an equilibrium configuration, with a singularity at the center, but with no event horizon, as trapped surfaces fail to form as the collapse proceeds. Such a family of naked singularity static spacetimes was in fact alluded to by Tolman many years ago \cite{tolman}. 
The accretion disk properties as well as the shadow properties for these models was studied in detail in \cite{Joshi:2013dva},\cite{Shaikh:2018lcc},
and these were compared to the black hole spacetimes in order to
explore the key possible theoretical differences, and possible observational implications were explored and examined.
In the next section, we outline the key details of these models, whereas for a further discussion, we refer to the above references.

The paper is organized as follows. The next section (\ref{JMNspt}) discusses the relevant properties of JMN spacetimes needed for our purpose here, of studying the timelike trajectories.
 In section (\ref{veff}), we discuss the effective potentials for timelike geodesics in JMN spacetimes and Schwarzschild spacetime, and also we compare the orbit equations for massive particles in these two geometries. In section (\ref{shapeorbit}), we start from an approximate solution of these orbit equations to explain various characteristics of particles orbits in the Schwarzschild and JMN spacetimes. This comparison of orbits brings out different distinguishable properties of particle orbits in the Schwarzschild black hole spacetime and JMN naked singularity spacetimes. We then conclude with a discussion of the final results and the possible future pointers.

\section{JMN Spacetimes}
\label{JMNspt}
The JMN-1 and JMN-2 spacetimes have the following line elements,
\begin{widetext}
\begin{eqnarray} \label{eq:Tmunu_effective}
 ds^2_{JMN-1} &=& -(1- M_0) \left(\frac{r}{R_b}\right)^\frac{M_0}{(1- M_0)}dt^2 + \frac{dr^2}{(1 - M_0)} + r^2d\Omega^2\,\, , 
\label{JMN-1metric} \\
 ds^2_{JMN-2} &=& -\frac{1}{16\lambda^2(2-\lambda^2)}\left[(1+\lambda)^2\left(\frac{r}{R_b}\right)^{1-\lambda}-(1-\lambda)^2\left(\frac{r}{R_b}\right)^{1+\lambda}\right]^2dt^2 + (2-\lambda^2)dr^2 + r^2d\Omega^2\,\, ,
 \label{JMN2metric}
\end{eqnarray}
\end{widetext}
where both the spacetimes match with the Schwarzschild 
geometry at the boundary $r=R_b$. In JMN-1, the positive constant $M_0$ should be always less than 1, and in JMN-2, $\lambda$ is  
a positive constant and its value is less than unity. 
Here, $d\Omega^2$ is $d\theta^2+\sin^2\theta d\phi^2$ due to
the spherical symmetry of JMN spacetimes. In this paper, we use the units where Newton's gravitational constant ($G_N$) and light velocity ($c$) is equal to one. 

Generally, modelling of a compact object is performed by considering a high density compact region in vacuum. Therefore, the exterior spacetime of a compact object is considered as the Schwarzschild spacetime. JMN spacetimes also can be glued with Schwarzschild spacetime at a boundary $r=R_b$. Therefore, we have a spacetime structure which is internally JMN-1 or JMN-2 and externally matched to a Schwarzschild spacetime, where the Schwarzschild geometry 
can be written as,
\begin{equation}
ds^2 = -\left(1-\frac{M_0R_b}{r}\right)dt^2 + \frac{dr^2}{\left(1-\frac{M_0R_b}{r}\right)} +r^2d\Omega^2\,\, , 
\label{schext}
\end{equation}
where the Schwarzschild radius ($R_s$) is $R_s=M_0R_b$, 
and as $0<M_0<1$, so we have $R_b>R_s$. The total Schwarzschild mass ($M_{TOT}$) is $M_{TOT}=\frac{M_0R_b}{2}$. In general relativity, for the matching of two spacetimes at a particular spacelike or timelike hypersurface, we need to satisfy two junction conditions: 

\begin{enumerate}

    \item The induced metrics of internal and external spacetimes on the matching hypersurface should be identical with each other. One can see that this is satisfied for the above two metrices (\ref{JMN-1metric}), (\ref{JMN2metric}) at the timelike hypersurface $r=R_b$. From the induced metric matching we get, $M_0=\frac{1-\lambda^2}{2-\lambda^2}$, for JMN-2.
    
    \item Another condition is to match the extrinsic curvatures ($K_{ab}$) at the hypersurface. Extrinsic curvature can be expressed in terms of the covariant derivative of normal vectors on the hypersurface: $K_{ab}=e^{\alpha}_ae^{\beta}_{b}\nabla_{\alpha}\eta_{\beta}$, where $e^{\alpha}_a$ is the tangent vectors on the hypersurface and $\eta^{\beta}$ is the normal to that hypersurface. 
\end{enumerate}

One can show that due to the zero radial pressure of JMN-1 spacetime, the extrinsic curvatures of JMN-1 and Schwarzschild spacetimes at $r=R_b$ are automatically matched with each other \cite{Dey:20192}. 
On the other hand, JMN-2 spacetime is seeded by an isotropic fluid, therefore, it can be shown that in order to match extrinsic curvature, the pressure in JMN-2 should vanish at $r=R_b$. Pressure in JMN-2 can be written as,
\begin{equation}
    P_{JMN-2} = \frac{1}{(2-\lambda^2)}\frac{1}{r^2}\left[\frac{(1-\lambda)^2 A-(1+\lambda)^2 Br^{2\lambda}}{A-Br^{2\lambda}}\right]\,\, ,
\end{equation}
where $ A = \frac{(1+\lambda)^2R_b^{\lambda-1}}{4\lambda\sqrt{2-\lambda^2}}$ and $B =  \frac{(1-\lambda)^2R_b^{-\lambda-1}}{4\lambda\sqrt{2-\lambda^2}}$. It can be checked that $P_{JMN-2}$ becomes zero at $r=R_b$.

As we mentioned, JMN spacetimes can be formed as an end state of gravitational collapse. It can be shown how a general collapsing metric with non-zero pressure can reach to an equilibrium state in asymptotic time without forming an apparent horizon. It may be worth noting here that, using Newtonian mechanics, one can show how a self-gravitating fluid with $N$ number of particles reaches to a virialized state in asymptotic time. Such a system of gravitationally interacting particles is considered to be virialized when, 
\begin{equation}
\langle T\rangle=-\frac{1}{2} \langle V_T \rangle,
\end{equation}
where $\langle T\rangle$ and $\langle V_T \rangle$ are the average kinetic energy and average total potential energy (over a large system time period) respectively. 
Analogous to Newtonian virialization process, the equilibrium process which is mentioned in 
\cite{Joshi:2011zm}, 
also takes an asymptotic time to equilibrate the collapsing matter cloud. This type of quasi-static collapse would be of relevance
in the context of dark matter halo formation and galaxy formation
as well \cite{Dey:20191},\cite{Dey:20192}. 
Therefore, the final asymptotic spacetimes (e.g. JMN-1, JMN-2 or others), which can be formed as an end state of the pressure supported quasi-static collapse may give some distinguishable astrophysical signatures between the black hole and naked singularity collapse final states. This is in the sense that since these spacetimes are static, non-vacuum solutions of Einstein equations, their timelike orbits should be distinguishable from the timelike orbits in the vacuum black hole spacetime, and it is these differences that are studied and
reported in this paper.

\section{The Particle orbit equations in Schwarzschild, JMN-1 and JMN-2 spacetimes }\label{veff}
\subsection{Timelike Geodesics in General Spherically Symmetric, Static Spacetime }
We can write a spherically symmetric, static spacetime as,
\begin{equation}
    ds^2 = - g_{tt}(r)dt^2 + g_{rr}(r)dr^2 + r^2(d\theta^2 + sin^2\theta d\phi^2)\,\, ,
    \label{static}
\end{equation}
where $g_{tt}$,$g_{rr}$ are the functions of $r$ only, and the azimuthal part of the spacetime shows the spherical symmetry. For a particle which is freely falling in this type of spacetime, the angular momentum ($h$) and the energy ($\gamma$) per unit of particle's rest mass are always conserved, where the angular momentum and energy conservation are the direct consequence of spherical and temporal symmetry of the above spacetime (\ref{static}). The conserved $h$ and $\gamma$ of a freely falling particle in the static, spherically symmetric spacetime can be written as,
\begin{eqnarray}
    h = r^2\frac{d\phi}{d\tau}\,\, ,\,\,\,
   \gamma =  g_{tt}(r)\frac{dt}{d\tau}\,\, ,
   \label{congen}
\end{eqnarray}
where $\tau$ is the proper time of the particle and we consider $\theta=\frac\pi2$. We know that freely falling particles always follow timelike geodesics, for which $v_{\mu}v^{\mu}=-1$, where $v^{\mu}$ is the particle's four-velocity. From this normalization of four-velocity we can write an effective potential which plays a crucial role on particles' trajectories in a spacetime. For the static and spherically symmetric spacetime which is mentioned in eq.~(\ref{static}), the effective potential can be written as,
\begin{equation}
V_{eff} = \frac{1}{2}\left[g_{tt}(r)\left(1+\frac{h^2}{r^2}\right)-1\right]
\label{veffgen}\,\, ,
\end{equation}
where the total energy ($E$) can be written as,
\begin{equation}
    E=\frac{g_{rr}(r)g_{tt}(r)}{2}\left(\frac{dr}{d\tau}\right)^2+V_{eff}(r)\,\, ,
    \label{totalE}
\end{equation}
where $E=\frac{\gamma^2-1}{2}$.
To find the stable circular trajectories of particles we need $V_{eff}(r_c)=E$, $V_{eff}^{\prime}(r_c)=0$ and $V_{eff}^{\prime\prime}(r_c)>0$, where $r_c$ is the radius of the circular orbit. If we change $h$ or $\gamma$ keeping all the other parameter of the metric constant, then the radius of the stable circular orbit also changes. Using $V_{eff}(r)=E$, $V_{eff}^{\prime}(r)=0$, one can write down the required $h$ and $\gamma$ for a circular orbit at a given radius,
\begin{equation}
    h^2=\frac{r^2}{\left(\frac{2g_{tt}(r)}{rg_{tt}^{\prime}(r)}-1\right)}\, ,\,\,\, \gamma^2=\frac{2g_{tt}^2(r)}{rg_{tt}^{\prime}(r)}\frac{1}{\left(\frac{2g_{tt}(r)}{rg_{tt}^{\prime}(r)}-1\right)}\,\, ,
    \label{he}
\end{equation}
\begin{figure}[t]
\includegraphics[scale=0.55]{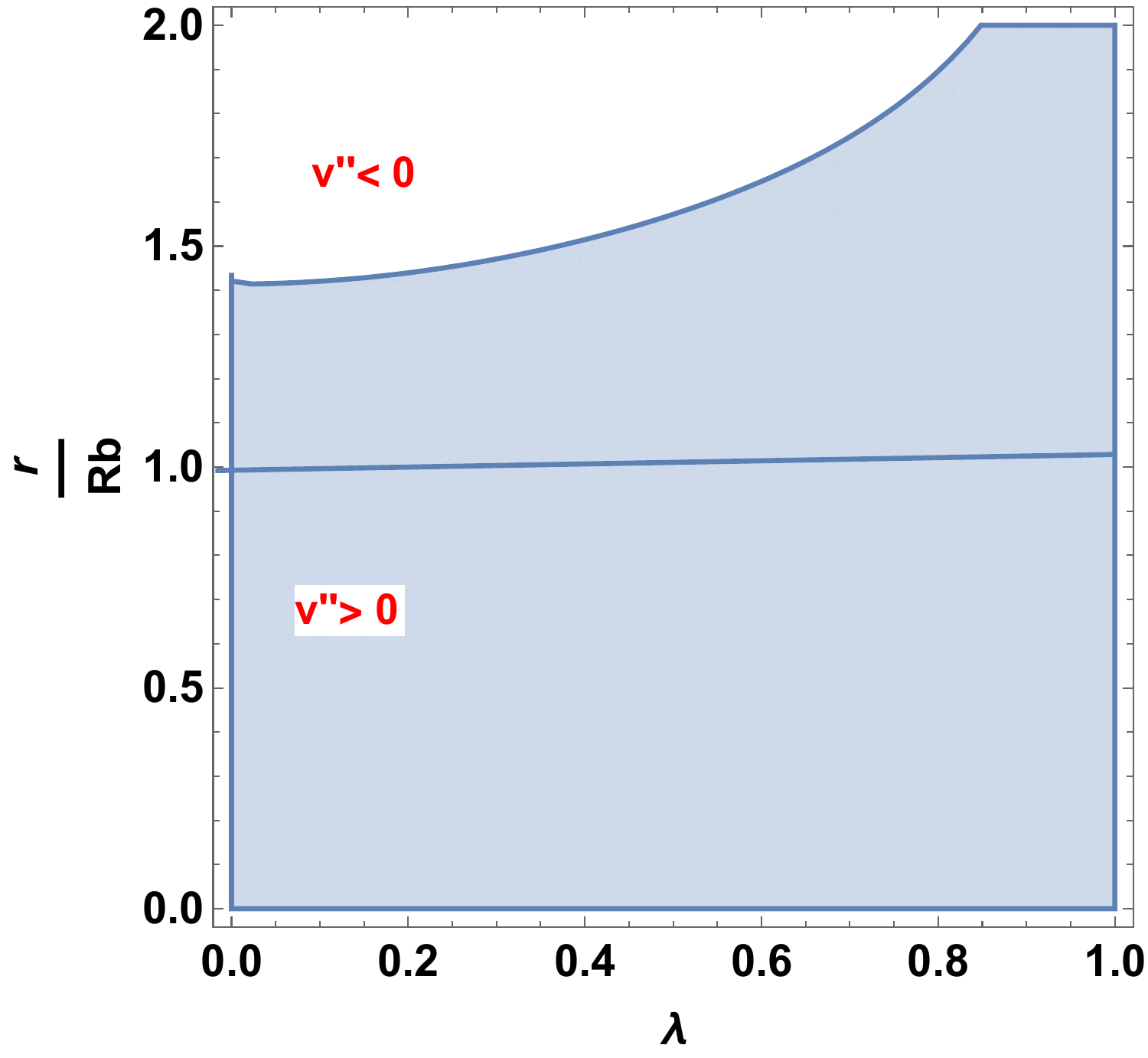}
\caption{Here the region where $(V^{\prime\prime}_{eff})_{JMN-2}>0$ is shown, where we have taken $R_b=1000$. We can see that $(V^{\prime\prime}_{eff})_{JMN-2}>0$ is always valid inside the region $0\leq\lambda<1$ and $0\leq r\leq R_b$.}
\label{vppjmn2}
\end{figure}
where the above conditions are true for both stable and unstable circular orbits. For stable circular orbits we have to consider another constrain $V_{eff}^{\prime\prime}(r)>0$. Bound non-circular orbits can be describe by solving $V_{eff}(r)=E$ which gives the radius of minimum approach ($r_{min}$) and maximum approach ($r_{max}$) of the particle towards the center. Therefore, one can define bound orbits of the freely falling particles in the following way,
\begin{eqnarray}
   V_{eff}(r_{min})=V_{eff}(r_{max})=E\, , \,\, \nonumber\\
   E-V_{eff}(r)>0\, ,\,\,\,\forall r\in (r_{min},r_{max}).
   \label{bound}
\end{eqnarray}
Now if we want to describe the shape of an orbit of a particle with some conserved value of $h$ and $E$, we then need to describe 
how the radial coordinate $r$ changes with azimuthal 
coordinate $\phi$. Using eq.~(\ref{totalE}) we get,
\begin{equation}
    \frac{d\phi}{dr}=\frac{h}{r^2}\frac{\sqrt{g_{rr}(r)g_{tt}(r)}}{\sqrt{2(E-V_{eff})}}
\end{equation}
For some simplification, the above equation can be written in terms of $u=\frac{1}{r}$, and then one can write a second order differential equation by differentiating the above equation with respect to $\phi$,
\begin{widetext}
 \begin{equation}
  \frac{d^2u}{d\phi^2}+\frac{u}{g_{rr}(u)}-\left[\frac{u^2}{2g_{rr}^2(u)}+\frac{1}{2g_{rr}^2(u)h^2}-\frac{\gamma^2}{2g_{tt}(u)g_{rr}^2(u)h^2}\right] \frac{dg_{rr}(u)}{du}+\frac{\gamma^2}{2g_{tt}^2(u)g_{rr}(u)h^2}\frac{dg_{tt}(u)}{du}  = 0\,\, .
  \label{orbiteqgen}
    \end{equation}
\end{widetext}
From the above orbit equation, it can be seen that the shape of the orbit depends totally on the spacetime where the test particle is freely falling.
\subsection{Orbit equations for Schwarzschild, JMN-1 and JMN-2 Spacetime}
While the Schwarzschild spacetime is a static, spherically symmetric, vacuum solution of Einstein equations, it has necessarily a strong singularity at the center which is covered by a null surface, the event horizon. As we stated in previous subsection, due to the temporal and spherical symmetry, Schwarzschild spacetime also has two conserved quantities: the energy ($\gamma_{SCH}$) and angular momentum ($h_{SCH}$) per unit rest mass of the freely falling particle in this spacetime. Using eq.~(\ref{congen}), the conserved quantities $\gamma_{SCH}$ and $h_{SCH}$ for Schwarzschild spacetime can be written as,
\begin{eqnarray}
   \gamma_{SCH} &=& \left(1-\frac{M_0R_b}{r}\right) \left(\frac{dt}{d\tau} \right)\,\, ,\\
 h_{SCH}&=& r^2  \left(\frac{d\phi}{d\tau}\right)\,\,,
 \end{eqnarray}
Using the general expression of the effective potential in the eq.~(\ref{veffgen}), the effective potential for Schwarzschild spacetime can be written as,
\begin{equation}
    (V_{eff})_{SCH}= \frac{1}{2}\left[\left(1 -\frac{M_0R_b}{r}\right)\left(1 + \frac{h_{SCH}^2}{r^2}\right) - 1\right]\,\, .
    \label{veffsch}
\end{equation}
From the above expression of effective potential, we can derive the expression of $h_{SCH}$ and $\gamma_{SCH}$ for circular geodesics,
\begin{eqnarray}
   \gamma_{SCH}^2=\frac{2\left(r-M_0R_b\right)^2}{r\left(2r-3M_0R_b\right)}\,,\, h_{SCH}^2=\frac{M_0R_br^2}{\left(2r-3M_0R_b\right)}\,\, ,\label{hesch}
\end{eqnarray}
where we use the general expressions of $\gamma$ and $h$ in eq.~(\ref{he}). From the above expression of conserved quantities for circular geodesics, it can be seen that no circular orbit is possible in the range: $0\leq r\leq \frac{3M_0R_b}{2}$. In terms of the total mass ($M_{TOT}$) of the black hole, this range can be written as $0\leq r\leq 3M_{TOT}$, where $M_{TOT}=\frac{M_0R_b}{2}$. However, for stable circular orbits, along with above two conditions in eq.~(\ref{hesch}), we need $ (V_{eff}^{\prime\prime})_{SCH}>0$. One can write the expression of $ (V_{eff}^{\prime\prime})_{SCH}$ as,
\begin{eqnarray}
   (V_{eff}^{\prime\prime})_{SCH}&=& \frac{2M_{TOT}}{r}\left(\frac{6M_{TOT}-r}{3M_{TOT}-r}\right)\,\, .
   \label{isco}
\end{eqnarray}
Therefore, for stable circular orbits we need $\gamma$ and $h$ as stated in eq.~(\ref{hesch}) and $r\geq6M_{TOT}$. There cannot be any stable circular orbits below $r=6M_{TOT}$ in Schwarzschild spacetime, therefore, $r=6M_{TOT}$ is known as the Innermost Stable Circular Orbit (ISCO). For bound orbits in this spacetime, the conditions which are stated in eq.~(\ref{bound}), should be fulfilled. Using the eq.~(\ref{orbiteqgen}), we can write the shape of particle's orbits in Schwarzschild spacetime,
\begin{equation}
   \frac{d^2u}{d\phi^2} + u = \frac{3M_0R_b}{2}u^2 + \frac{M_0R_b}{2h^2}\,\, .
   \label{orbiteqsch}
\end{equation}
\begin{figure*}
\centering
\subfigure[particle orbits in Schwarzschild where $M_0 =0.09$,$h=200$,$E=-0.0268$]
{\includegraphics[scale=0.5]{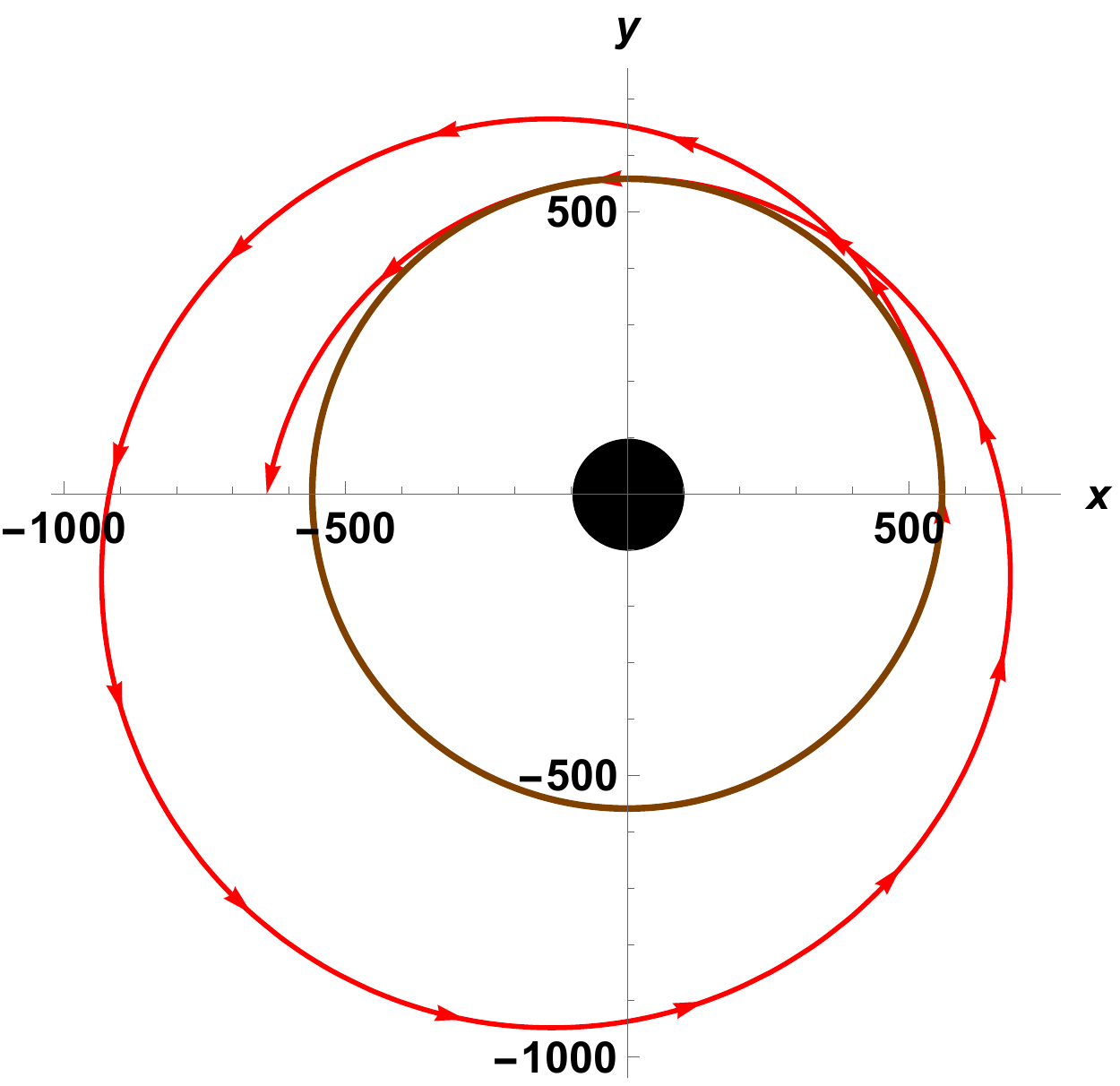}}
\hspace{0.2cm}
\subfigure[Particle orbit in jmn-1 where $M_0 =0.09$,$h=200$, $E=-0.0268$]
{\includegraphics[scale=0.5]{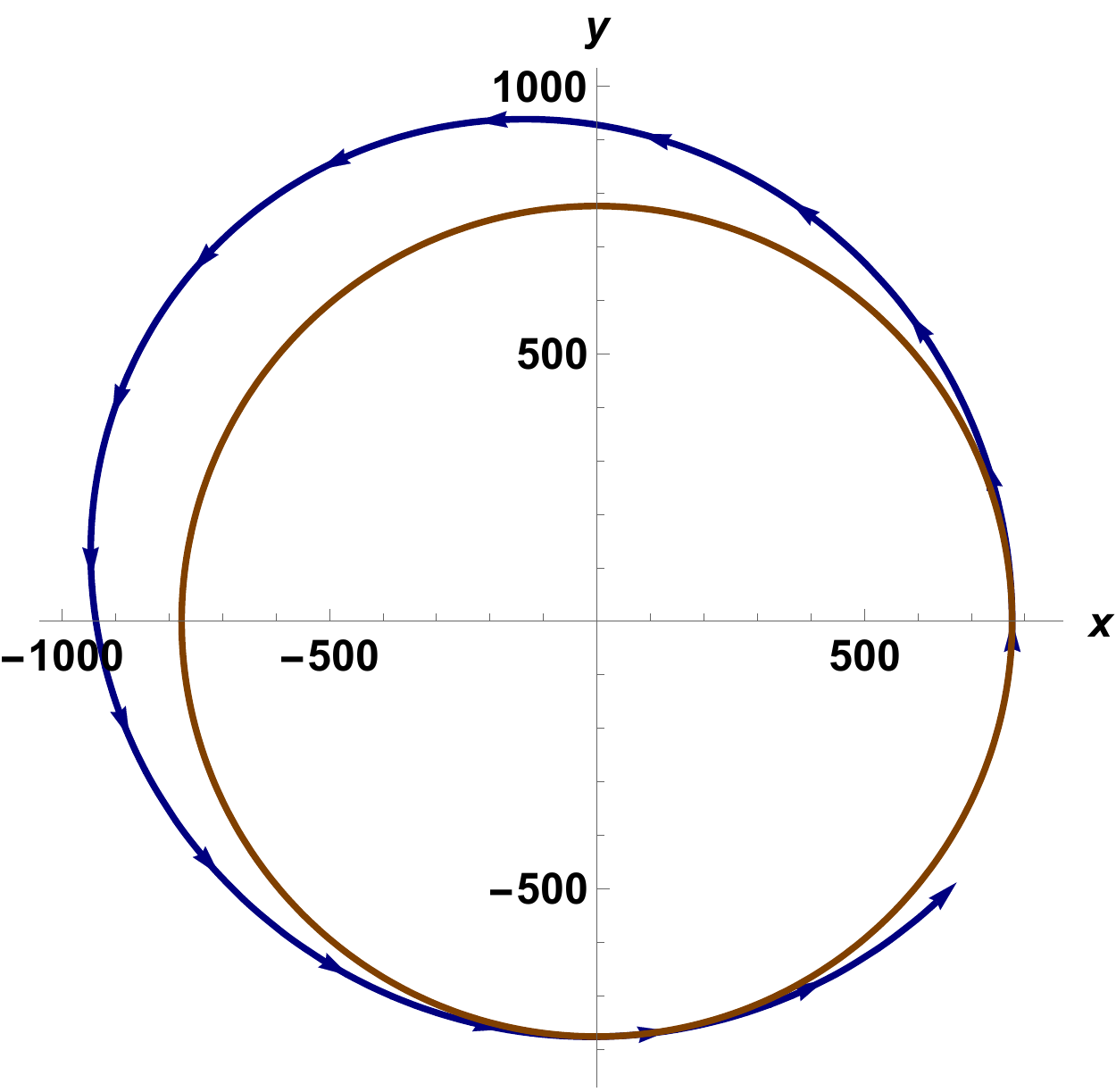}}
\hspace{0.2cm}   
\subfigure[Particle orbit in jmn-1 where $M_0 =0.55$,$h=1100$, $E=-0.004$]
{\includegraphics[scale=0.5]{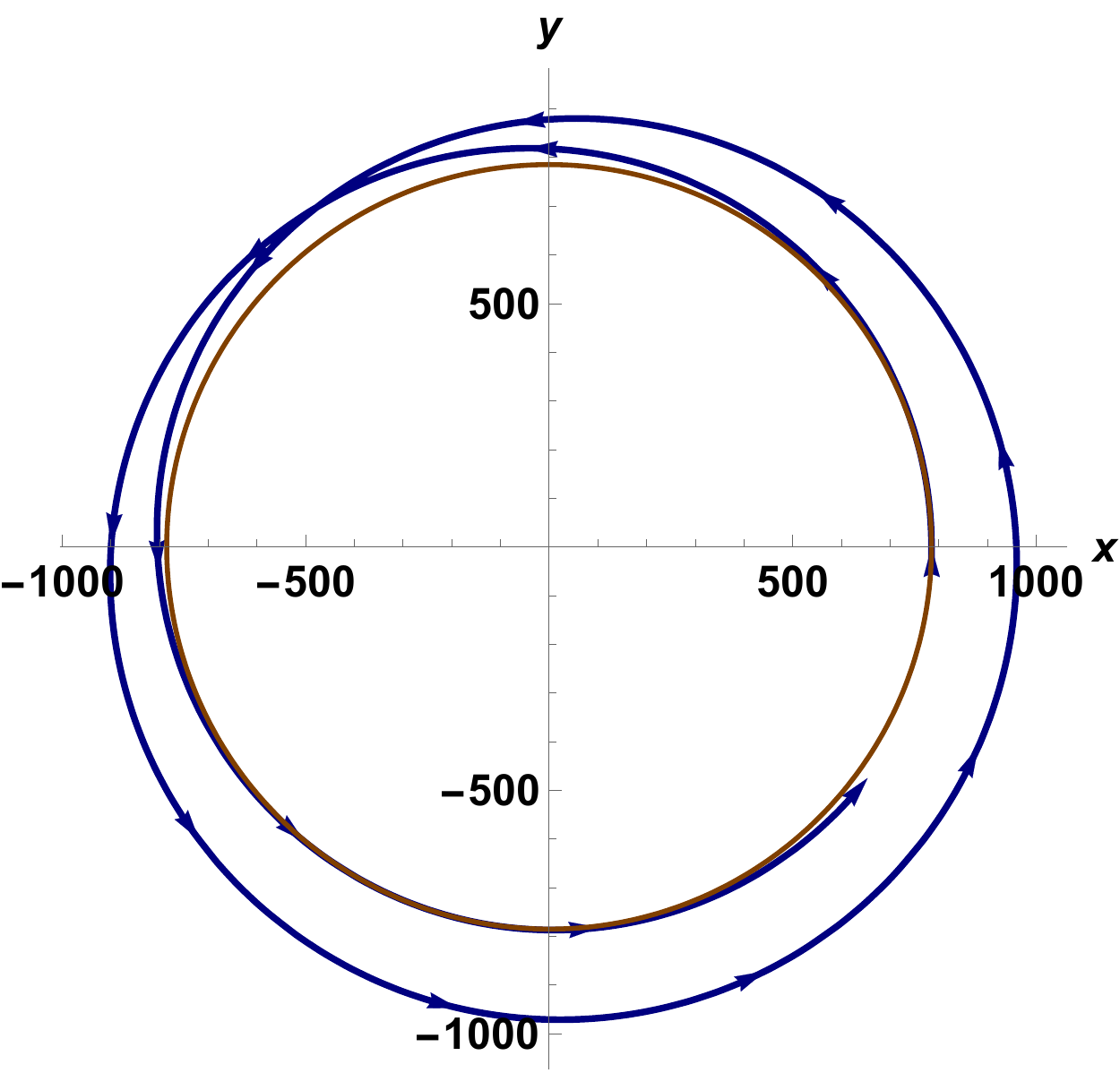}}
\hspace{0.2cm}
\subfigure[Particle orbit in JMN-2 where $\lambda=0.9$, $h=290$, $E=-0.0445$]
{\includegraphics[scale=0.5]{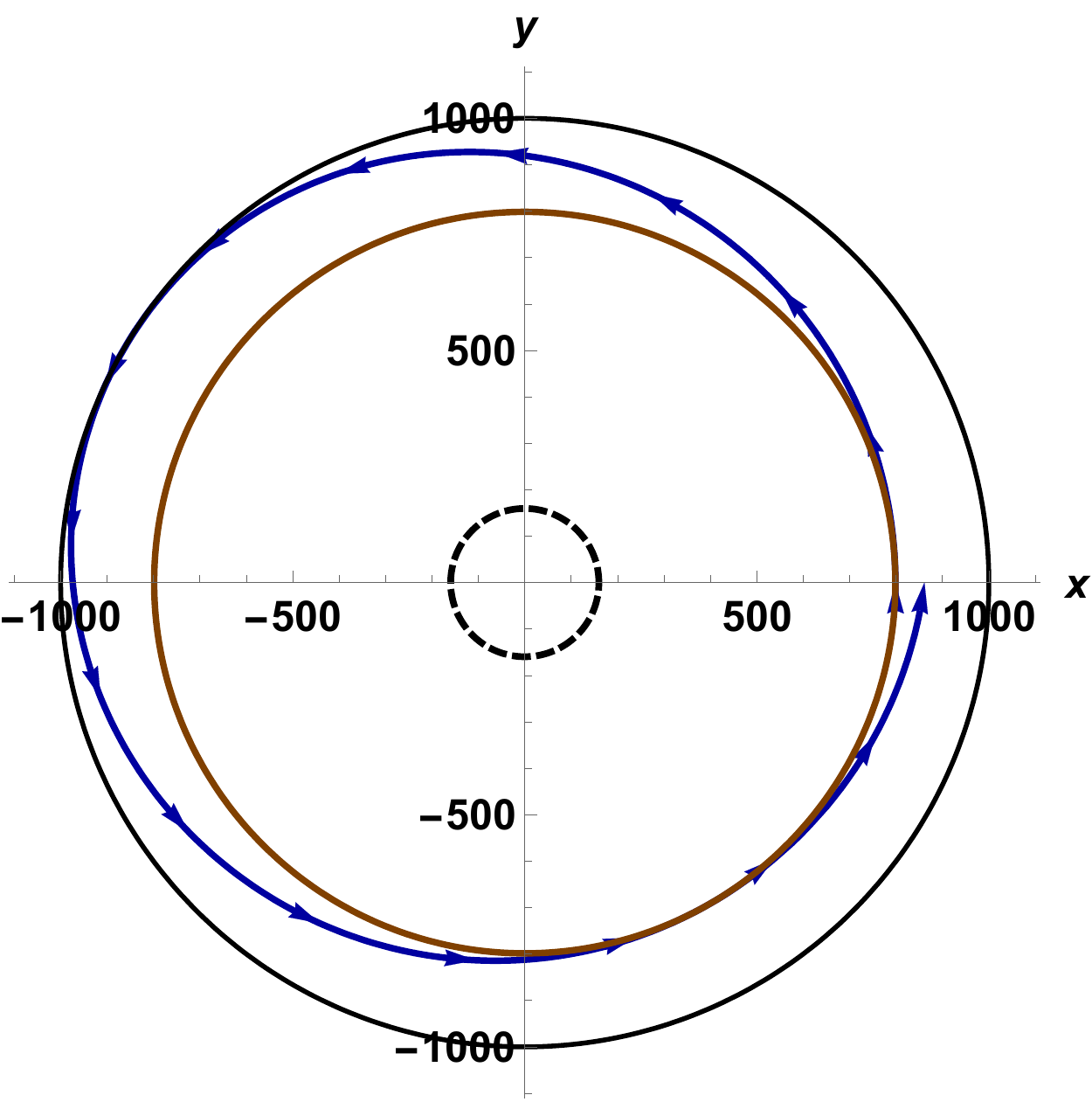}}
\hspace{0.2cm}
 \subfigure[Particle orbit in JMN-2 where $\lambda=0.1$, $h=892$, $E=-0.04885$]
 {\includegraphics*[scale=0.5]{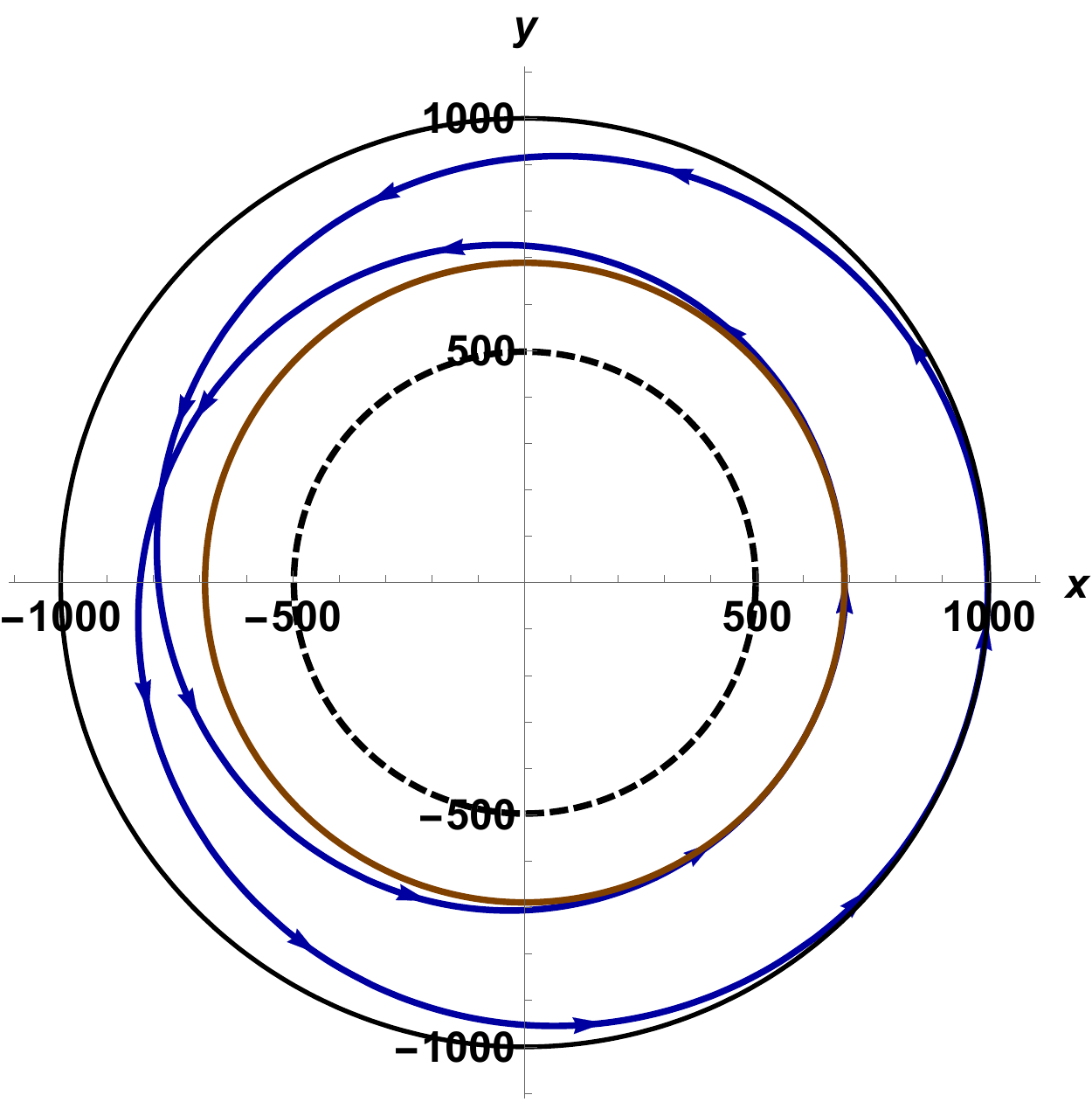}}
 \caption{In this figure, particle orbits in Schwarzschild, JMN-1 and JMN-2 are shown. It can be seen that for $M_0=0.09$ (JMN-1) and $\lambda=0.9$ (JMN-2), the angular distance travelled by the particle to reach one perihelion point to another perihelion point is less than $2\pi$, whereas for  $M_0=0.55$ (JMN-1) and $\lambda=0.1$ (JMN-2) it reaches after $2\pi$ rotation.  The black dark region in the diagram of orbit in Schwarzschild spacetime shows the position of the black hole and the brown circle shows the minimum approach of the particle towards the center. The black circle outside the orbit in JMN-2, shows the position of matching radius $R_b=1000$.      }
 \label{precgen}
\end{figure*}
As we mentioned before, JMN-1 and JMN-2 spacetimes are spherically symmetric, static, non-vacuum solutions of Einstein equations. These spacetimes are seeded by anisotropic fluid (with zero radial pressure) and isotropic fluid respectively. Both these spacetimes can be formed as an asymptotic end state of a quasi-static gravitational collapse process \cite{Joshi:2011zm},\cite{Joshi:2013dva}. Both these spacetimes have strong curvature singularities at the center. There are no event horizons in these spacetimes, therefore, the quantum gravity effects around the singularities at the center can be visible by asymptotic observer, at least in principle. Using previously discussed technique, we can derive now the effective potential for JMN-1 and JMN-2 spacetimes,
\begin{widetext}
\begin{eqnarray}
(V_{eff})_{JMN-1} &=& \frac{1}{2}\left[(1- M_0)\left(\frac{r}{R_b}\right)^\frac{M_0}{(1- M_0)}\left(1 + \frac{h_{JMN-1}^2}{r^2}\right) - 1\right]\,\, ,\\
(V_{eff})_{JMN-2} &=& \frac{1}{2}\left[\frac{1}{16\lambda^2(2-\lambda^2)}\left[(1+\lambda)^2\left(\frac{r}{R_b}\right)^{1-\lambda}-(1-\lambda)^2\left(\frac{r}{R_b}\right)^{1+\lambda}\right]^2\left(1 + \frac{h_{JMN-2}^2}{r^2}\right) - 1\right]\,\, ,
\end{eqnarray}
\end{widetext}
where $h_{JMN-1}$ and $h_{JMN-2}$ are the conserved angular momentum per unit rest mass of the test particles in JMN-1 and JMN-2 spacetimes respectively. For JMN-1 spacetime, the conditions for stable circular orbits are,
\begin{eqnarray}
   \gamma^2_{JMN-1}&=&\frac{2(1-M_0)^2\left(\frac{r}{R_b}\right)^{\frac{M_0}{1-M_0}}}{\left(2-3M_0\right)}\, , \, \\
   h^2_{JMN-1}&=&\frac{r^2M_0}{2-3M_0}\,\, ,\\
   (V_{eff}^{\prime\prime})_{JMN-1}&=&\frac{M_0}{R_b^2}\left(\frac{r}{R_b}\right)^{\frac{3M_0-2}{1-M_0}}>0\,\, .
\end{eqnarray}
From the above three equations, it can be understood that for $M_0<\frac23$, JMN-1 spacetime has stable circular orbits of any radius. Therefore, unlike Schwarzschild spacetime, in JMN-1 spacetime, for $M_0<\frac23$, there is no such region where unstable circular orbits exist. On the other hand, for $M_0>\frac23$ there is no solution for stable circular orbits at any value of $r$, as $h_{JMN-1}$ and $\gamma_{JMN-1}$ become imaginary.
As we previously mentioned, the $\lambda$ parameter in JMN-2 space-time should be inside the range $0\leq \lambda<1$ and $M_0=\frac{1-\lambda^2}{2-\lambda^2}$. Now, if we consider the matching radius $R_b=nM_{TOT}$ where $n$ is a positive number, we can write,
$R_b=n\frac{M_0R_b}{2}=\frac{nR_b(1-\lambda^2)}{2(2-\lambda^2)}$, which implies,
\begin{equation}
    \lambda=\sqrt{\frac{n-4}{n-2}}\,\, .
\end{equation}
The above equation shows that the matching radius $R_b$ for JMN-2 spacetime should be $4M_{TOT}\leq R_b<\infty$. One can see that when $n\rightarrow\infty$ or $R_b\rightarrow\infty$, the value of $\lambda\rightarrow 1$. Therefore, $\lambda$ close to unity implies a largely extended compact object which has JMN-2 geometry inside, and outside the object, the spacetime is Schwarzschild. On the other hand, in JMN-1 spacetime there is no minimum limit of $R_b$, $R_b$ can have any value between zero to infinity. Using eq.~(\ref{he}), we can write the conditions for circular orbits in JMN-2 spacetime as,
\begin{widetext}
\begin{eqnarray}
 \gamma_{jmn-2}^2 &=& \frac{1}{16}\left[\frac{r^2\left\lbrace-(1+\lambda)^2+(\lambda-1)^2\left(\frac{r}{R_b}\right)^{2\lambda}\right\rbrace ^3\left(\frac{r}{R_b}\right)^{-2\lambda}}{\lambda^3(\lambda^2-2)\left\lbrace(1+\lambda)^2+(\lambda-1)^2\left(\frac{r}{R_b}\right)^{2\lambda}{2\lambda}\right\rbrace R_b^2}\right]\,\, ,\\
h_{jmn-2}^2 &=& \left[\frac{r^2(1-\lambda)(\lambda+1)\left\lbrace(1+\lambda)-(1-\lambda)\left(\frac{r}{R_b}\right)^{2\lambda}\right\rbrace}{\lambda\left[(\lambda+1)^2+(\lambda-1)^2\left(\frac{r}{R_b}\right)^{2\lambda}\right]}\right]\,\, .
\end{eqnarray}
\end{widetext}
We can see from the above expressions of $\gamma_{JMN-2}$ and $h_{JMN-2}$ that for $0\leq\lambda<1$ we can get circular orbits. For stability of circular orbits, we need $(V^{\prime\prime}_{eff})_{JMN-2}>0$. Due to the large analytical expression of $(V^{\prime\prime}_{eff})_{JMN-2}$, in fig.~(\ref{vppjmn2}) we numerically show the region where it is greater than zero. One can see that $(V^{\prime\prime}_{eff})_{JMN-2}>0$ is possible for the region $0\leq\lambda<1$ and $0\leq r\leq R_b$. Therefore, like JMN-1 spacetime, JMN-2 spacetime also can have circular orbits very close to the center.  
Therefore, in both JMN-1 and JMN-2 space-times, particles can have stable circular orbit at any value of $r\leq R_b$. Using general orbit eq.~(\ref{orbiteqgen}) for spherically symmetric, static spacetime, one can write the following orbit equations of a particle freely falling in JMN-1 and JMN-2 spacetime respectively,
\begin{widetext}
\begin{eqnarray}
   \frac{d^2u}{d\phi^2} + (1 - M_o) u - \frac{\gamma^2}{2h^2}\frac{M_0}{(1- M_0)}\left(\frac{1}{u}\right)\left(\frac{1}{uR_b}\right)^\frac{-M_0}{(1- M_0)}&=&0\,\, ,\label{orbiteqJMN-1}\\
   \frac{d^2u}{d\phi^2} + \frac{u}{(2-\lambda^2)} - \frac{16\lambda^2 \gamma^2u^{1+2\lambda}}{h^2}\left[\frac{(1+\lambda)^2}{(R_b)^{1-\lambda}}(1-\lambda)u^{2\lambda} - \frac{(1-\lambda)^2}{(R_b)^{1+\lambda}}(1+\lambda)\right]\left[\frac{(1+\lambda)^2}{(R_b)^{1-\lambda}}u^{2\lambda} - \frac{(1-\lambda)^2}{(R_b)^{1+\lambda}}\right]^{-3}&=&0
   \label{orbiteqJMN2}
\end{eqnarray}
\end{widetext}
\begin{figure}
\centering
\subfigure[$M_0 =0.05$,$h=105.26$, $E=-0.02$, $R_b=1000$]
{\includegraphics[scale=0.58]{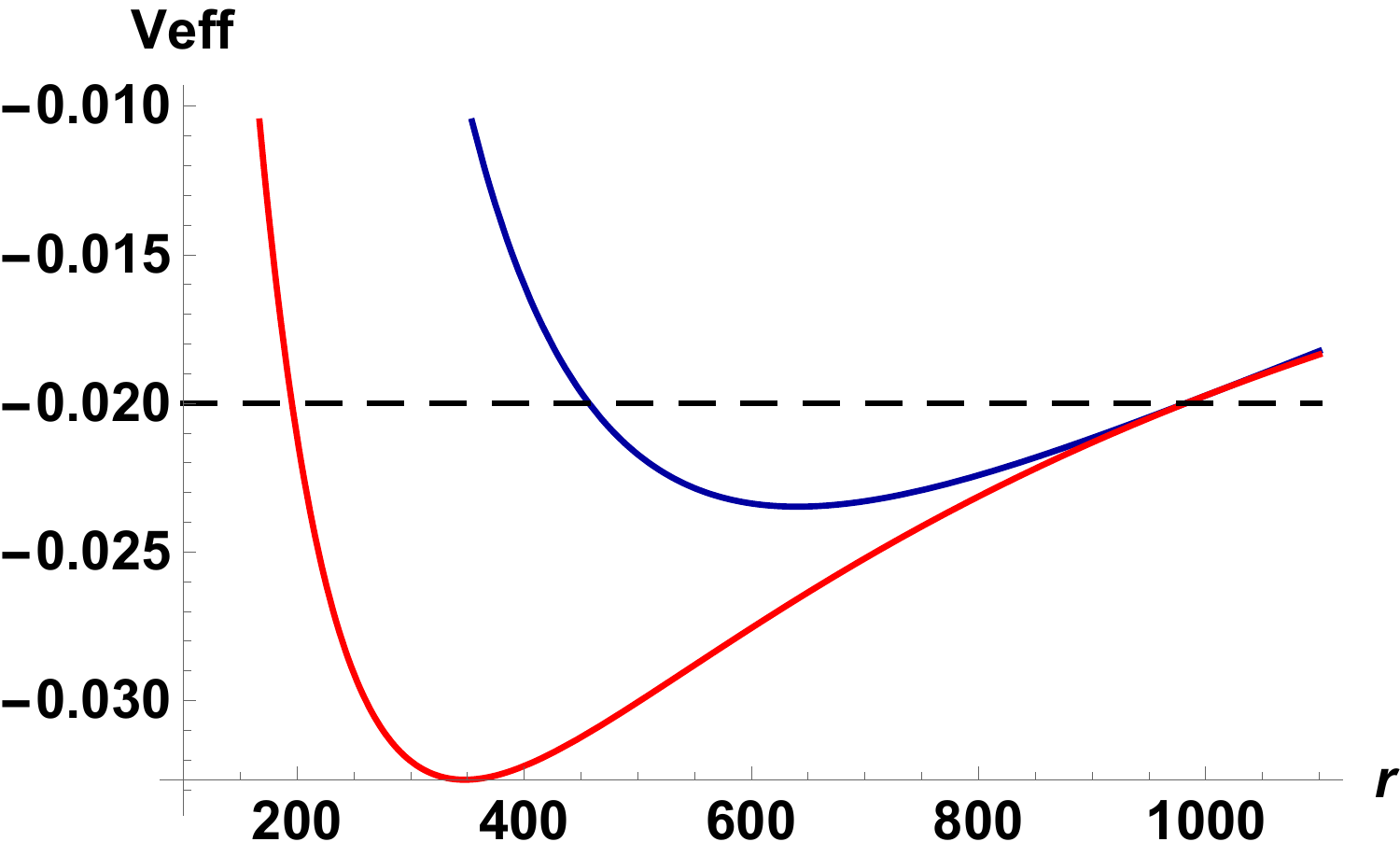}\label{Veffcompare1}}\\
\subfigure[Particle orbit in Schwarzschild spacetime]
{\includegraphics[scale=0.58]{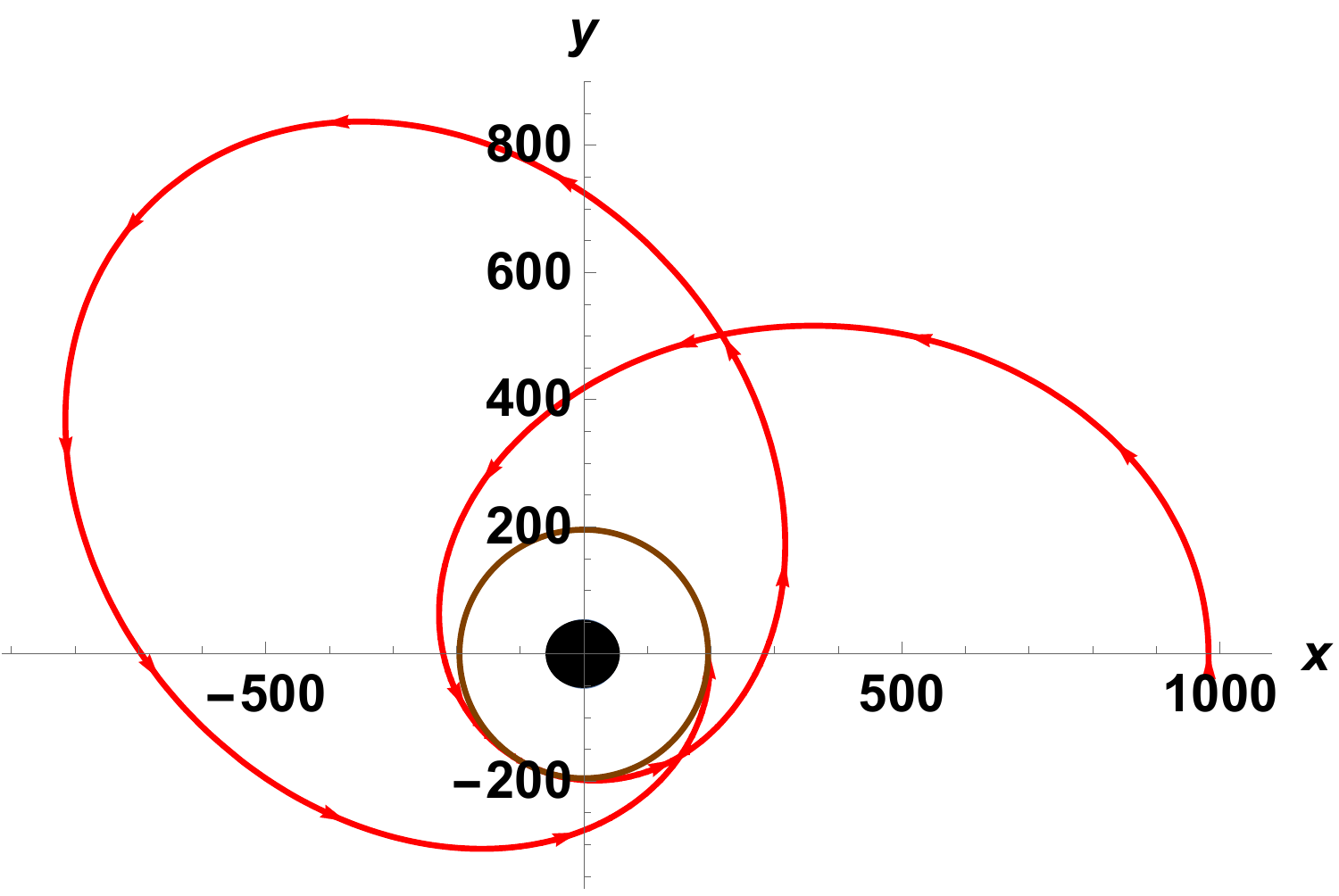}\label{sch-orbit1}}
\subfigure[Particle orbit in JMN-1 spacetime]{
   \includegraphics[scale=0.58]{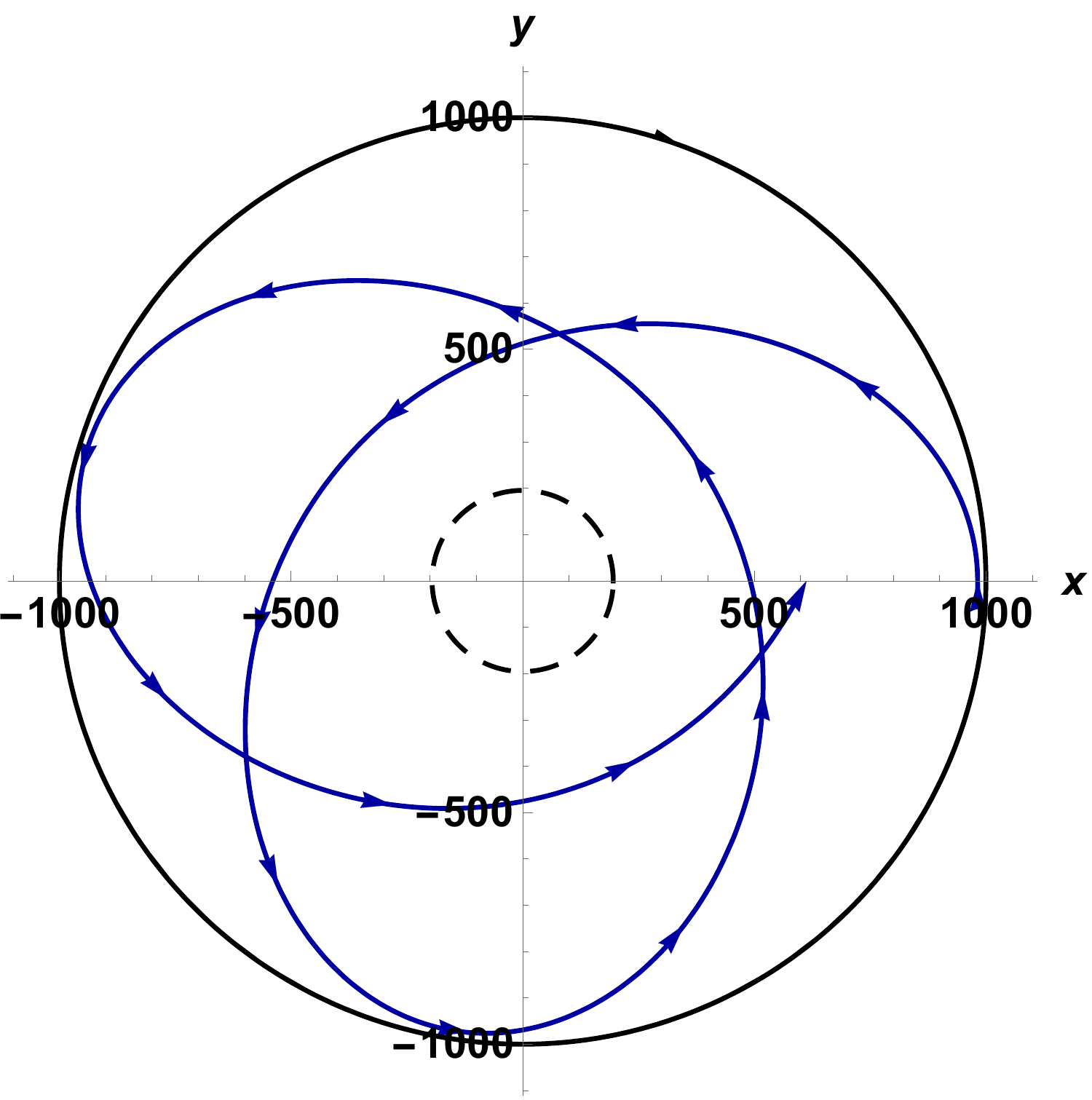}\label{JMN-orbit1}
 }
\caption{
Here, particle orbits in Schwarzschild and JMN-1 spacetimes are shown, where we take the Schwarzschild mass ($M_{TOT}=\frac{M_0R_b}{2}=25$), same for both spacetimes. In the effective potential diagram, the blue line and red line correspond to JMN-1 and Schwarzschild spacetime respectively and the dotted horizontal black line is indicating the total energy of the freely falling particle. The black circle outside the orbit in JMN-1, shows the position of matching radius $R_b=1000$, which is greater than the Schwarzschild radius $R_s=M_0R_b=50$, which is shown by dotted black circle.}
\label{lessRb}
\end{figure}

\begin{figure*}
\centering
\subfigure[$M0 = 0.008, h=20, E=-0.003
$]
{\includegraphics[scale=0.85]{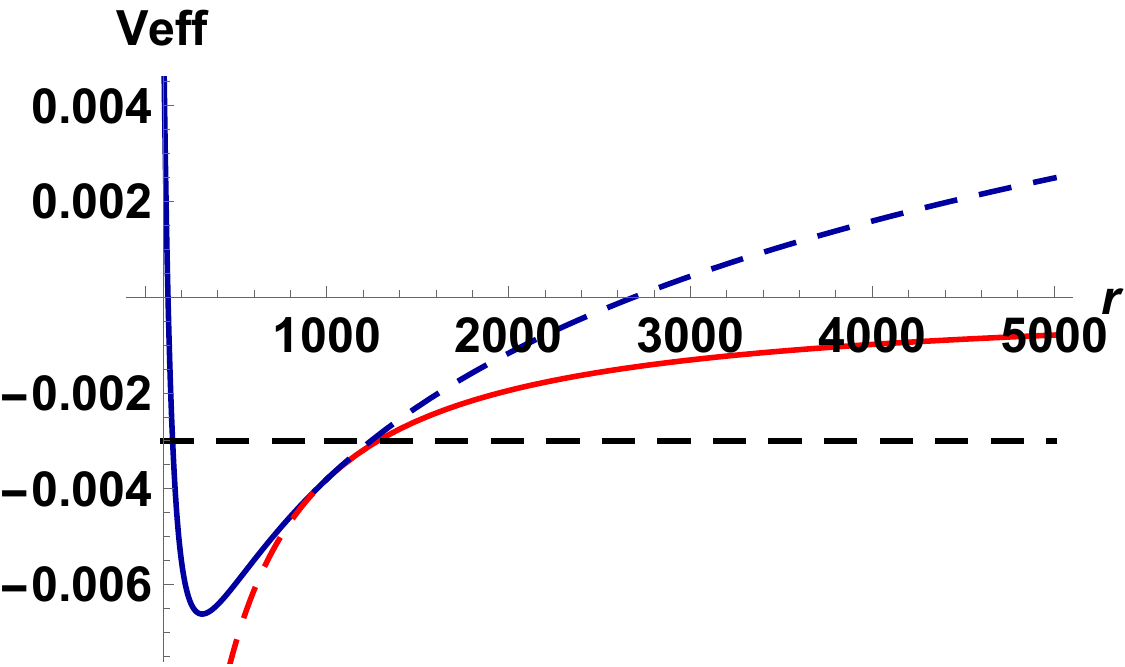}
\label{potperturb} }
\\
 \subfigure[Particle orbit in Schwarzschild spacetime]{
   \includegraphics[scale=0.82]{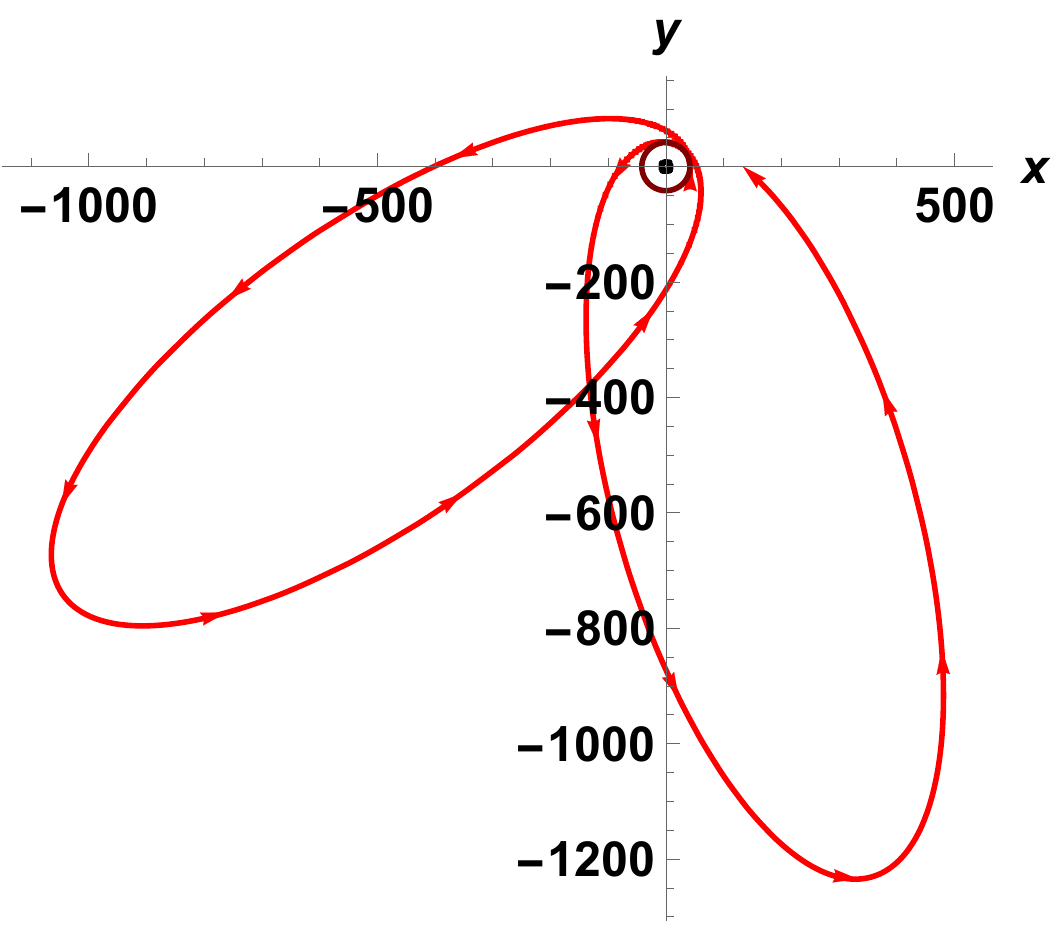}
\label{orbitsch22} }\hspace{0.1cm}
 \subfigure[Particle orbit partly in JMN-1 and partly in Schwarzschild spacetime.  ]{
   \includegraphics[scale=0.82]{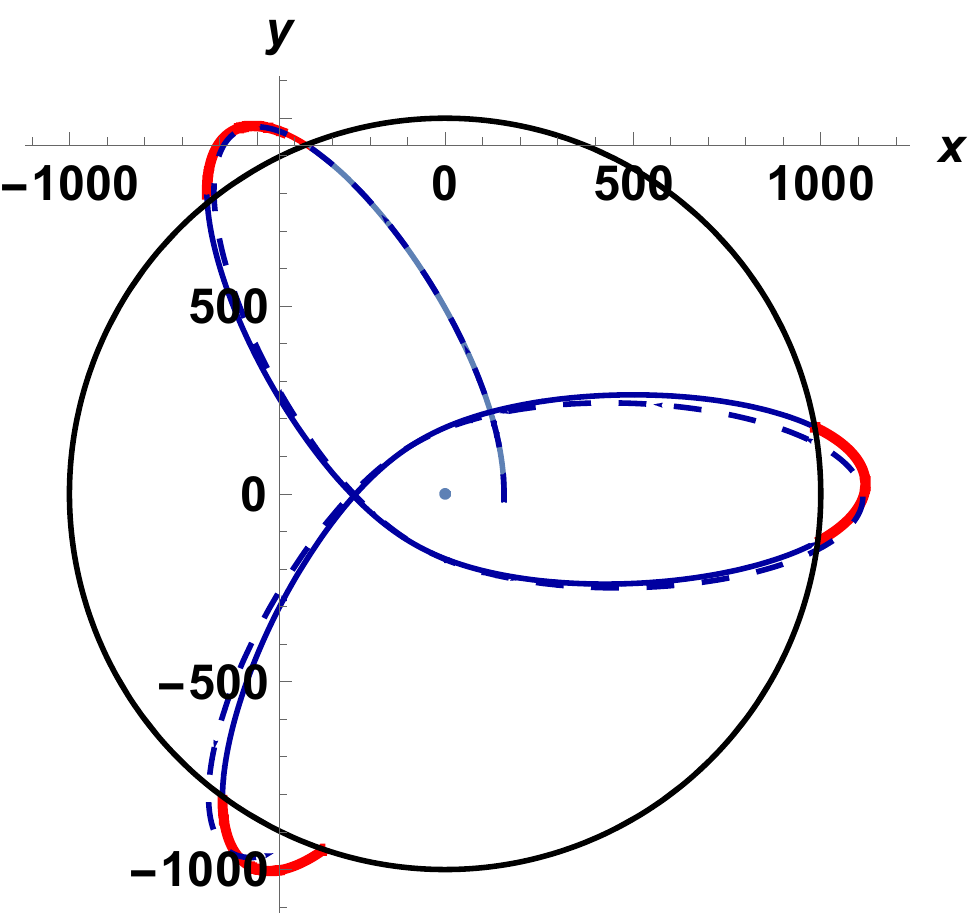}\label{orbitJMN-1grrb}
 }
 \caption{ In this figure, we show the trajectory of a particle when it crosses the matching radius $R_b$. In the third diagram the dotted blue line shows what would be the particle trajectory if there is no Schwarzschild spacetime outside. On the other hand, the solid blue line shows particle's actual path in JMN-1 spacetime and the solid red line shows particle's trajectory in Schwarzschild spacetime. }
\label{greaterRb} 
\end{figure*}

\section{Shape of bound orbits in Schwarzschild and JMN spacetimes}\label{shapeorbit}
The orbit equations which are written in eq.~(\ref{orbiteqsch}), (\ref{orbiteqJMN-1}) and (\ref{orbiteqJMN2}) can give us the information about the shape of different orbits of freely falling particles in Schwarzschild, JMN-1 and JMN2 spacetimes respectively. One can numerically solve those orbit equations, however, if we obtain the approximation solution of those orbit equations, then that  can be used to understand some important properties of the orbits in a better way. As we know, the Schwarzschild spacetime is asymptotically flat, therefore, the orbit eq.~(\ref{orbiteqsch}) should tends to Newtonian orbit equation in weak field limit. One can see that in asymptotic limit eq.~(\ref{orbiteqsch}) becomes Newtonian orbit equation,
$$\frac{d^2u}{d\phi^2}+u=\frac{M_0R_b}{2h^2}\,\, ,$$ where we neglect the $u^3$ term as an asymptotic limit. As we know, the solution of the above equation is $u=\frac{M_0R_b}{2h^2}\left[1+e \cos(\phi)\right]$, which suggests that, in asymptotic limit, the orbiting particle will reach its previous position after one full $2\pi$ rotation. However, the same is not true when the particle is near the center and we cannot neglect the $u^3$ term in the orbit equation in the corresponding Schwarzschild spacetime. 

If we consider the case when the eccentricity of a particle's orbit is small enough to neglect the second and higher order contributions of eccentricity ($e$), we can write the following approximate solution for the orbit eq.~(\ref{orbiteqsch}) of Schwarzschild spacetime,
\begin{equation}
u=\frac{1}{p}\left[1+e\cos(m\phi)+O(e^2)\right]\,\, ,
\label{orbitsch1}
\end{equation}
\begin{figure*}
\centering
\subfigure[$M0 = 0.62, h=1250, E=-0.02
$]{
   \includegraphics[scale=0.68]{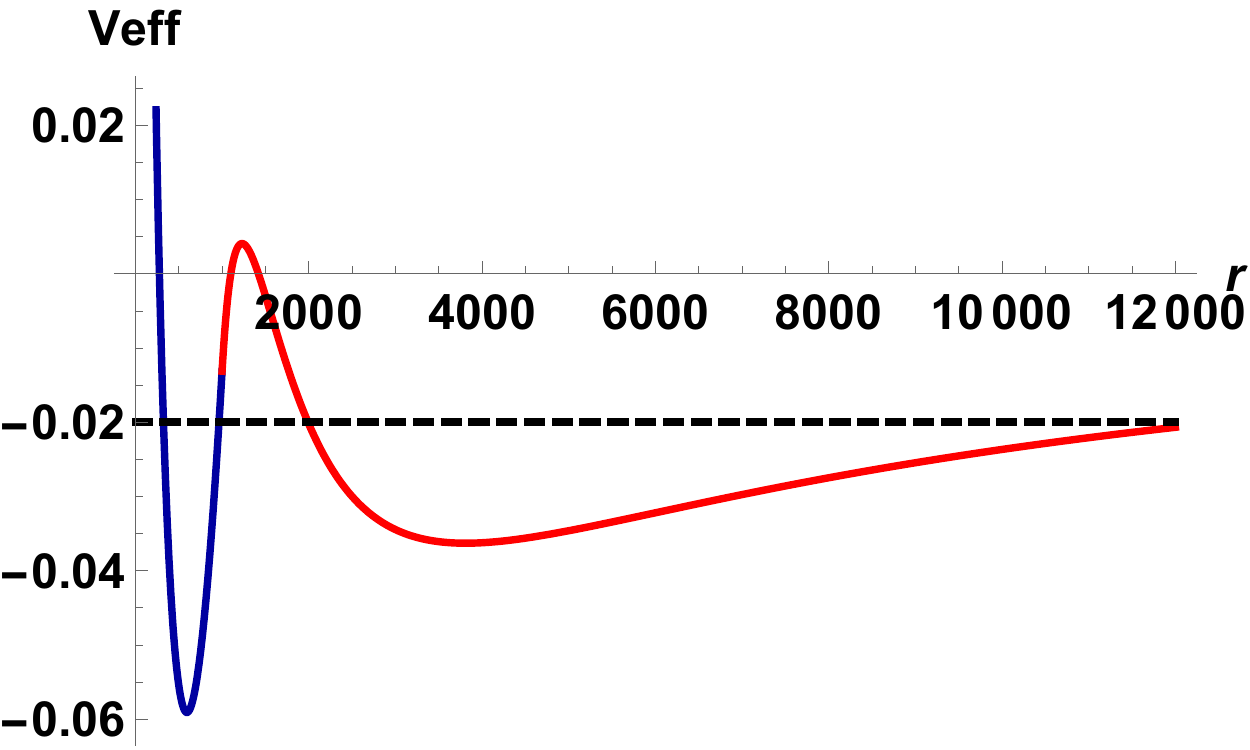}\label{Veffcompare3}
 }
 \subfigure[Particle orbits in Schwarzschild spacetime]{
   \includegraphics[scale=0.65]{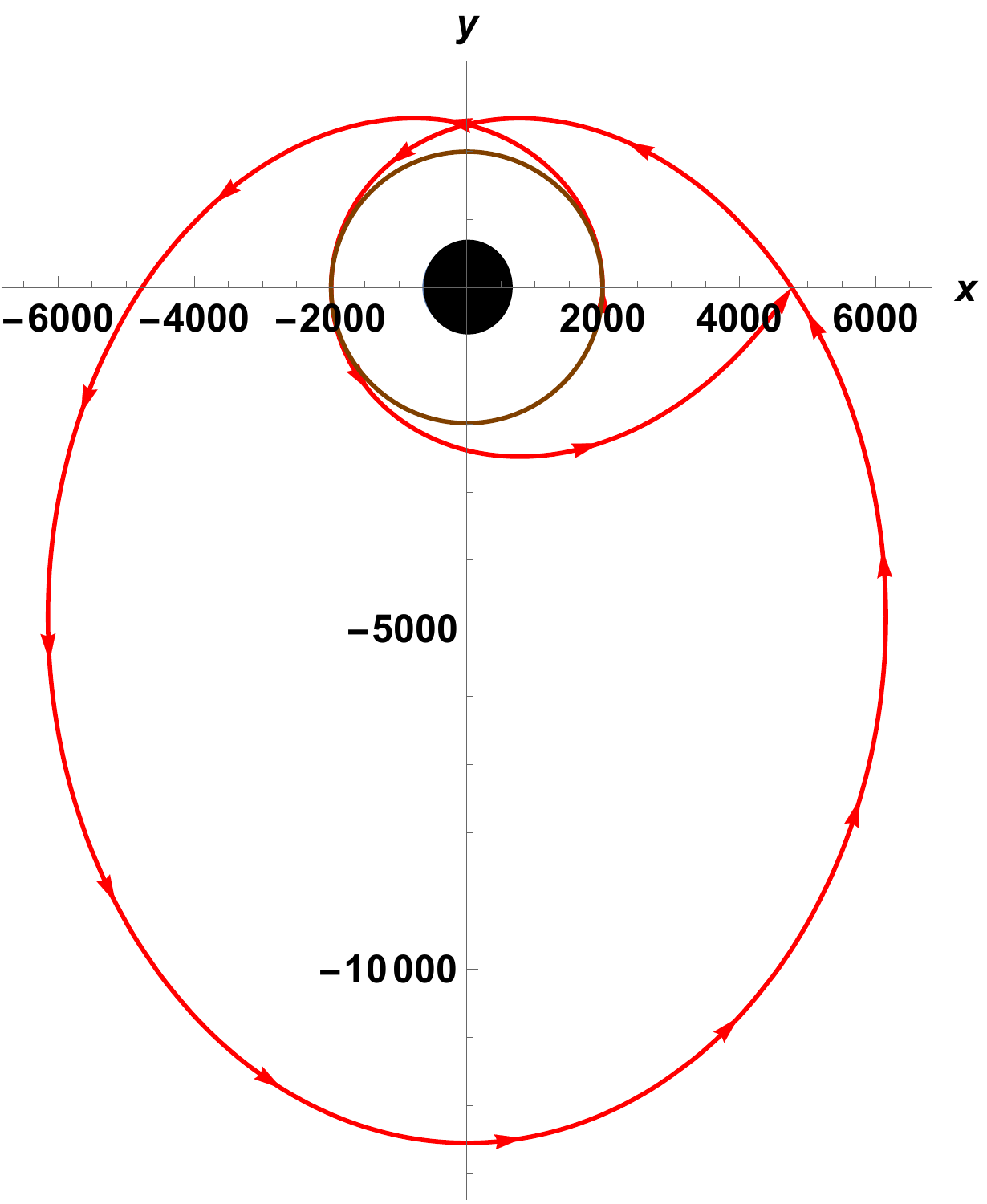}
\label{orbitsch2} 
}\\
\subfigure[Particle orbits in JMN-1 spacetime]{
   \includegraphics[scale=0.64]{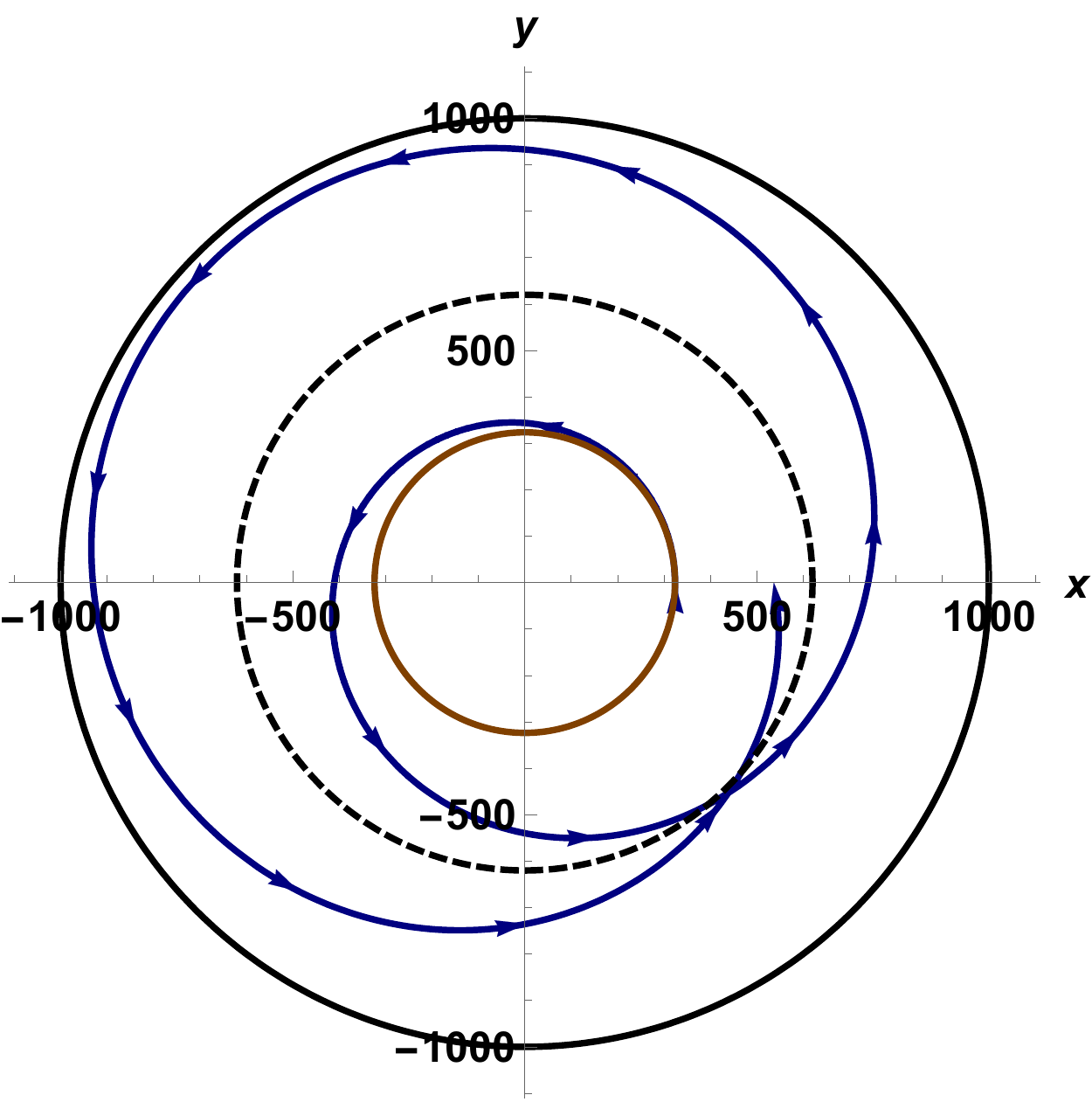}\label{orbitJMN-13}
 }
 \caption{In this diagram, we take $M_0=0.62$. As $M_0>0.6$, the minimum value of effective potential in Schwarzschild spacetime is always outside the matching radius $R_b$. In the effective potential diagram, the blue line is for JMN-1 spacetime, whereas the red line corresponds to Schwarzschild spacetime. The blue line ends at the matching radius $R_b=1000$.   }
 \label{M062}
\end{figure*}
where $m$ and $p$ are positive real numbers. As the orbit equation in Schwarzschild spacetime asymptotically approaches to Newtonian orbit equation, one can write the above form of solution. Using eq.~(\ref{orbitsch1}),(\ref{orbiteqsch}), one can get the following expression of $p$ and $m$ by neglecting second order and higher order terms,
\begin{eqnarray}
   p=\frac{1+\sqrt{1-\frac{3M_0^2R_b^2}{h^2}}}{\frac{M_0R_b^2}{h^2}}\,\, ,\label{psch}\\
   m=\sqrt{1-\frac{3M_0}{p}}\label{msch}\,\, ,
\end{eqnarray}
where we do not write the another solution of $p=\frac{1-\sqrt{1-\frac{3M_0^2R_b^2}{h^2}}}{\frac{M_0R_b^2}{h^2}}$, as $m$ becomes imaginary when $h>\sqrt3M_0R_b$, which is a necessary condition for real values of $p$. When $h>\sqrt3M_0R_b$, we always have $p>3M_0$ and $0<m<1$, which implies that starting from the closest approach to the center (perihelion position), after a full $2\pi$ rotation particle would not reach its previous perihelion position. To reach the perihelion position again, the particle needs some extra angle of rotation which depends upon the value of $m$. This precession of the closest approach of the orbit is known as perihelion precession. In Schwarzschild case, as $0<m<1$, the orbit precesses in the direction of particle rotation. In weak field limit, we can take $h\gg\sqrt3M_0R_b$, which implies $p=\frac{2h^2}{M_0R_b^2}$.and $p\gg3M_0$. With this limit, we can write the approximate expression of $m$ as, $m=\left(1-\frac{3M_0^2R_b^2}{4h^2}\right)$. Therefore, in weak field limit, if a particle starts from perihelion position, after a rotation of angle $\frac{2\pi}{\left(1-\frac{3M_0^2R_b^2}{4h^2}\right)}$ it will again reach the perihelion position. Therefore, the particle has to travel extra $$\delta\phi_{prec}=\frac{6\pi M_0^2R_b^2}{4h^2}=\frac{6\pi M_{TOT}^2}{h^2}\,\,.$$ This $\delta\phi_{prec}$ is the weak field limit of the precession angle of a particle's orbit in Schwarzschild spacetime. 

Coming now to the JMN case, 
the approximate solution of the orbit equation~(\ref{orbiteqJMN-1}) corresponding to the JMN-1 spacetime can also be written considering small value of eccentricity $e$. If we transform the $u\rightarrow uR_b$, we can write the eq.~(\ref{orbiteqJMN-1}) in the following form,
\begin{equation}
\tilde{u}\frac{d^2\tilde{u}}{d\phi^2}+\frac{1}{1+2\delta}\tilde{u}^2=C_{\delta}\tilde{u}^{2\delta}\,\, ,
\label{orbiteq2}
\end{equation}
\begin{figure*}
\centering
\subfigure[Effective potential for $Rb/h=0.1$]{
   \includegraphics[scale=0.65]{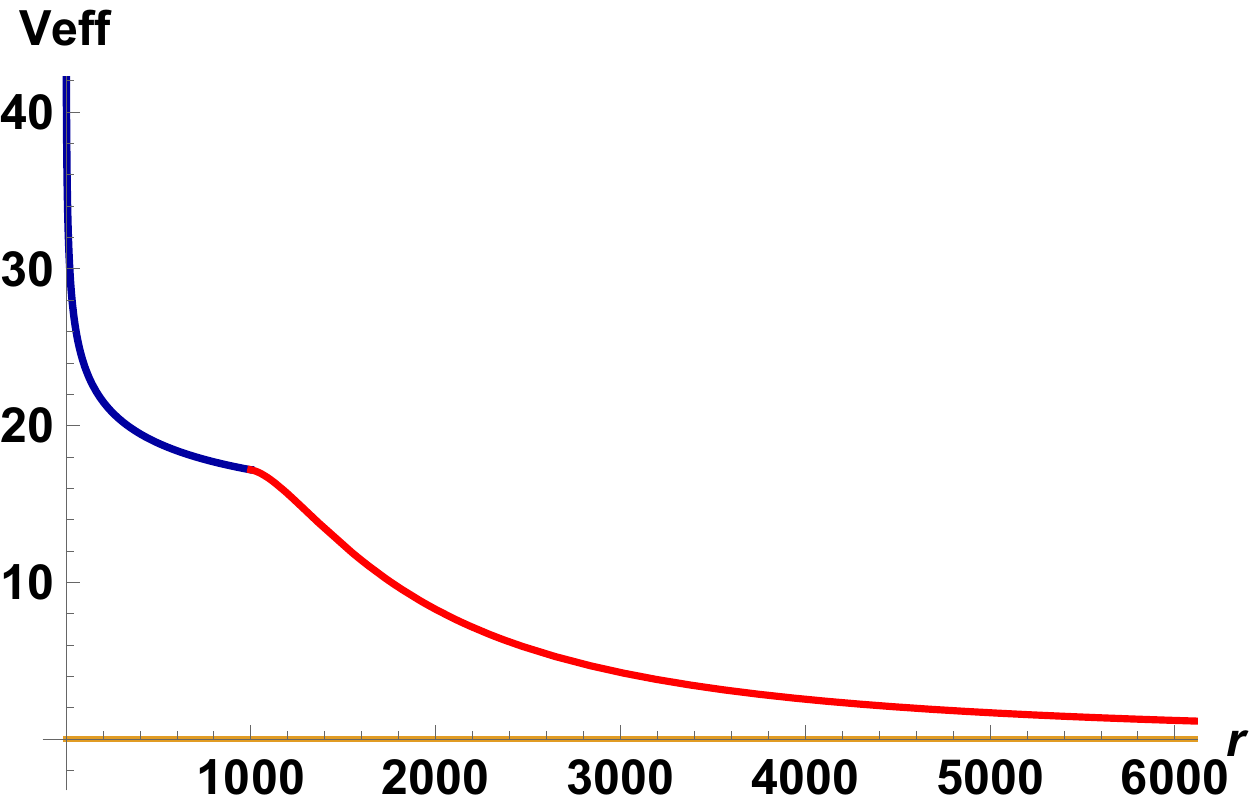}
 \label{bb11}}
\subfigure[Effective potential for $Rb/h=0.4$]{
   \includegraphics[scale=0.65]{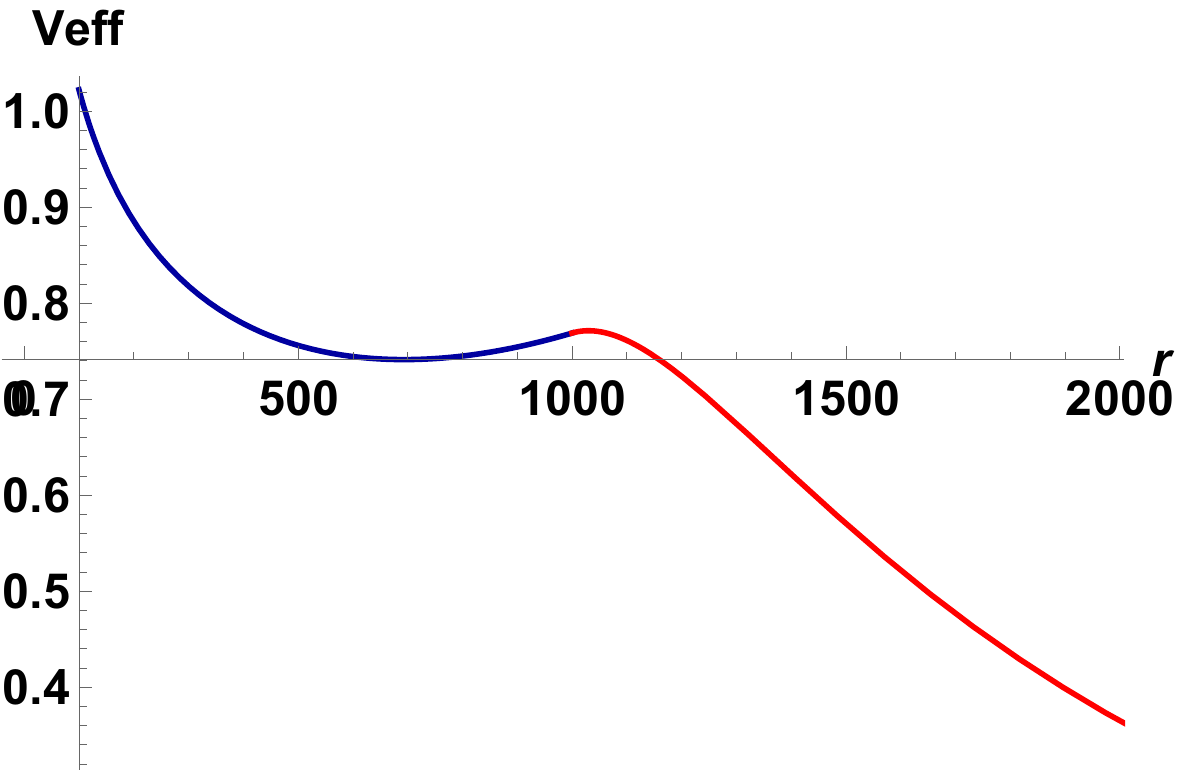}
\label{cc11} 
}
\\
\subfigure[Effective potential for $Rb/h=0.9$]{
   \includegraphics[scale=0.65]{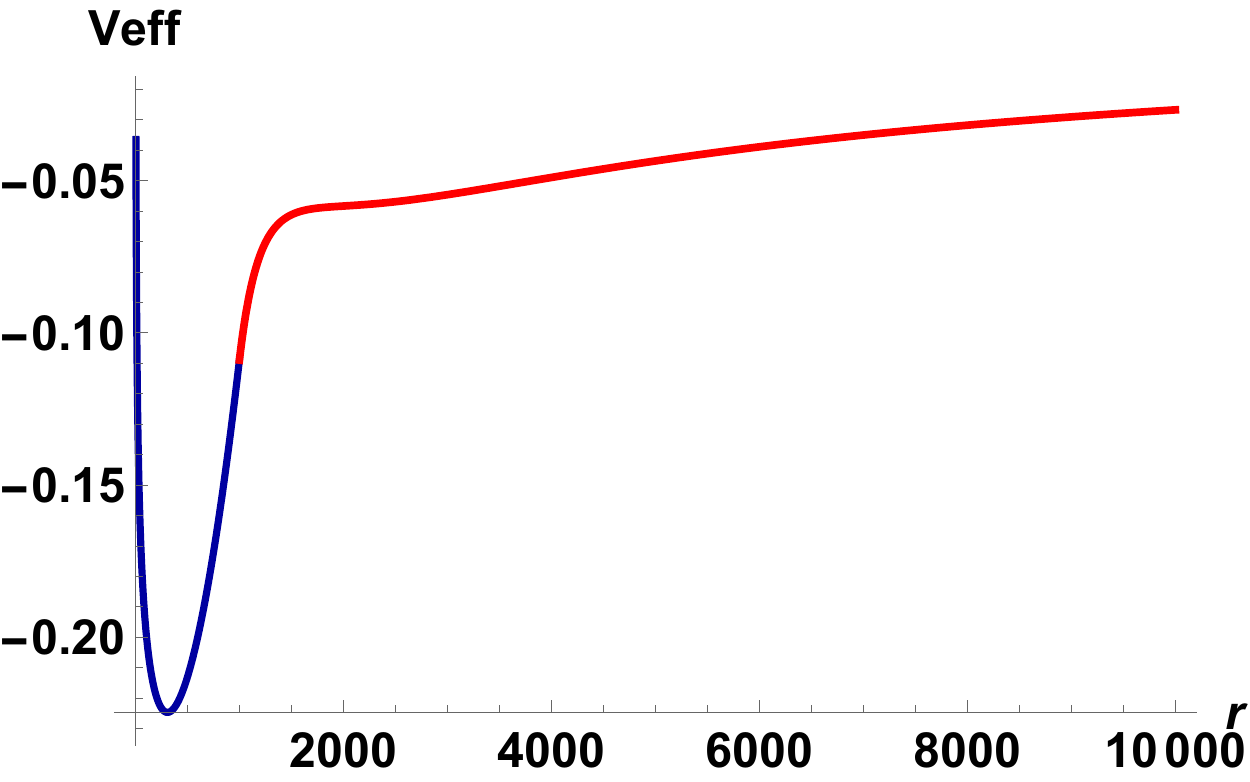}
 \label{aa11new}}
\caption{This diagram shows how the behaviour of effective potential changes with $h$, where we take $M_0 =0.65$ and $R_b=1000$.} 
\label{potbehav}
\end{figure*}
where $\delta=\frac{M_0}{2(1-M_0)}$, $C_{\delta}=\frac{\gamma^2R_b}{h^2}\delta$ and $\tilde{u}=uR_b$. 
In \cite{Struck:2005hi},\cite{Struck:2005oi}, 
the author introduces an approximate solution for this type of differential equations. The solution can be written as,
\begin{equation}
\tilde{u}=\frac1p\left[1+e\cos(m\phi)+O(e^2)\right]^{\frac12+\delta}\,\, .
\label{approxorbit}
\end{equation}
Using eq.~(\ref{orbiteqJMN-1}) and eq.~(\ref{approxorbit}) we can get the following expressions of $p$ and $m$,
\begin{eqnarray}
p&=&\left[C_{\delta}(1+2\delta)\right]^{-\frac{1}{2(1-\delta)}}\,\, ,\nonumber\\
m&=&\sqrt{\frac{2(1-\delta)}{2\delta+1}}=\sqrt{2-3M_0}\,\, .
\label{m}
\end{eqnarray}
From the above expression of $m$, it can be seen that for $0< M_0<\frac13$, $m$ is greater than one, whereas for $\frac13< M_0<\frac23$, $m$ is less than one. Therefore, in JMN-1 spacetime, starting from the perihelion position, a particle can reach the same position before (for $0< M_0<\frac13$ ) and after (for $\frac13< M_0<\frac23$) one $2\pi$ rotation. Consequently, the nature of the orbits also changes across $M_0=\frac13$. For $M_0<\frac13$, the orbit precesses in the opposite direction of particle motion, whereas, for $M_0>\frac13$ we get Schwarzschild like precession. In JMN-2, also we can have two type of precession.  In fig.~(\ref{precgen}), we show all these properties of orbits in JMN-1 and JMN-2 and compare it with the orbits in Schwarzschild spacetime. In that figure, it can be seen that the particle starts form its closest approach $r_c=558.941$ in Schwarzschild spacetime and again reaches to that radius after a $2\pi$ rotation. On the other hand, in that figure, it is shown that in JMN-1 and JMN-2 spacetimes, particle can reach the perihelion position before and after the $2\pi$ rotation.
It follows that the particle orbits and the precession behaviours are 
significantly different in these black hole and naked singularity spacetimes.

Using eq.~(\ref{m}), one can easily show that the orbits in JMN-1 spacetime have no precession when $M_0=\frac13$ or $m=1$. When $M_0$ is very small compared to $\frac23$, we can write the eq.~(\ref{orbiteq2}) as,
\begin{equation}
\tilde{u}\frac{d^2\tilde{u}}{d\phi^2}+\tilde{u}^2=C_\delta\,\, ,
\label{orbiteq3}
\end{equation}
where we consider the value of $R_b$ to be large, so that $C_{\delta}=\frac{\gamma^2R_b}{h^2}\delta$ cannot be neglected. In Newtonian mechanics, the above orbit equation appears when one takes a logarithmic potential. For  $0<M_0\ll\frac23$, $m$ approaches $\sqrt2$. Therefore, in this limit, in JMN-1 spacetime, particle reaches to the perihelion point after 254.56 degrees of rotation. 

As we previously mentioned, in this paper we pedagogically compare the bound orbits in Schwarzschild spacetimes and bound orbits in a spacetime structure where it is internally JMN-1 or JMN-2,   
externally matched to a Schwarzschild spacetime. We basically show how a freely falling particle, with a particular angular momentum and total energy, moves in Schwarzschild spacetime and in a naked singularity spacetime structure, where the Schwarzschild mass $M_{TOT}$ is same for both Schwarzschild spacetime and the other spacetime structure. In Fig.~(\ref{lessRb}), it is shown that the minimum value of effective potentials in Schwarzschild and JMN-1 spacetimes are inside the matching radius $R_b=1000$. It is possible to have particles' whole trajectory inside the matching radius $R_b$ when the following inequality holds,
\begin{equation}
    \frac{R_b}{h}>\sqrt{\frac{2-3M_0}{M_0}}\,\, .
\end{equation}
One can obtain the above relation by considering the minimum value of effective potential of JMN-1 spacetime inside the matching radius $R_b$. With the above inequality, a particle needs to have certain amount of total energy to be inside the interior JMN-1 spacetime. Now, maintaining the above condition, $\frac{R_b}{h}$ can have arbitrary large values which gives bound trajectories of particle inside the JMN-1 spacetime. However, as it was discussed previously (eq.~(\ref{psch})), in Schwarzschild spacetime there exists a lower limit of $h$ below which no bound timelike orbits are possible. For $h>\sqrt3M_0R_b$ or $\frac{R_b}{h}<\frac{1}{\sqrt3M_0}$, particles in Schwarzschild spacetime can have bound orbits. Therefore, to compare with bound orbits in Schwarzschild spacetime, we need to satisfy the following condition,
\begin{equation}
  \sqrt{\frac{2-3M_0}{M_0}}<\frac{R_b}{h}<\frac{1}{\sqrt{3}M_0}\,\, .
\label{ineq1}  
\end{equation}
If the above condition is satisfied, with some suitable total energy, a massive particle can have entire trajectory inside the interior JMN spacetime.
In fig.~(\ref{lessRb}), we take 
$M_0 =0.05$, therefore the range of $\frac{R_b}{h}$ is, $6.08<\frac{R_b}{h}<11.547$. To maintain this inequality of $\frac{R_b}{h}$, we consider  $h=105.26$ and $R_b=1000$, where $\frac{R_b}{h}=9.5$ and to get the entire particle trajectory inside the matching radius $R_b$, we take $E=-0.02$. These parameter values which satisfy the inequality of eq.~(\ref{ineq1}), allow us to compare the bound, timelike trajectories in a Schwarzschild spacetime with the bound, timelike trajectories in JMN-1 spacetime. With the above mentioned parameter values, it can be seen in the fig.~(\ref{sch-orbit1}) and fig.~(\ref{JMN-orbit1}) that the orbit precession in Schwarzschild spacetime is in the direction of particle motion, however, on the other hand, in JMN-1 spacetime particle's orbits precesses in the opposite direction of the particle's motion.

In fig.~(\ref{lessRb}), it can be seen that a particle with energy $E=-0.015$, cannot have whole bound orbit inside the matching radius $R_b=1000$. This situation is described in fig.~(\ref{greaterRb}), where we take $M0 = 0.008, h=20, E=-0.003$. With the total energy $E=-0.003$, one can see in fig.~(\ref{potperturb}) that the particle will cross the matching radius $R_b$. In fig.~(\ref{orbitJMN-1grrb}), we show that as particle crosses the matching radius, the trajectory (shown by solid blue and solid red line) of the particle changes from what it should be (shown by dotted blue line) in JMN-1 spacetime.     

Keeping the matching radius fixed ar $r=R_b$, if we make $h$ greater than the value $R_b\sqrt{\frac{M_0}{2-3M_0}}$, no bound orbit will be possible inside the interior JMN-1 spacetime, and only in the external Schwarzschild spacetime particles can have bound orbits. Therefore, there exists a maximum limit of angular momentum (per unit rest mass) $h$, below which particle's full trajectory can be inside the matching radius $R_b$. On the other hand, if $h< \sqrt{3}M_0R_b$, there will be no bound orbits in Schwarzschild spacetime.  As $ \sqrt{3}M_0R_b<R_b\sqrt{\frac{M_0}{2-3M_0}}$, it is possible that the particle can have full trajectories inside the matching radius. This type of situation is shown in the fig.~(\ref{aa11new}). Also in JMN-2 spacetime, we have the following inequality for which it is possible that the particle can have bound trajectories inside the JMN-2 spacetime and also in Schwarzschild spacetime,
\begin{equation}
    \sqrt{\frac{1+\lambda^2}{1-\lambda^2}}<\frac{R_b}{h}< \frac{2-\lambda^2}{\sqrt{3}\left(1-\lambda^2\right)}\,\, .
\end{equation}
An important point here is, as JMN spacetimes have no event horizon, particles with some finite angular momentum can go very close to the central strong singularity. On the other hand, as we know, in Schwarzschild spacetime, a particle having a bound orbit can only reach up to $r=4M_{TOT}=2M_0R_b$ and below $r=6M_{TOT}$, stable circular orbit is not possible. In JMN-1 spacetime, particle with $h< M_0R_b\sqrt{\frac{M_0}{2-3M_0}}$, can have bound trajectories inside the Schwarzschild radius.  
In the fig.~(\ref{M062}), with $M_0 = 0.62, h=1250, E=-0.02$, we show that particle orbits in JMN-1 spacetime can be inside the Schwarzschild radius $R_s$,
In fig.~(\ref{Veffcompare3}), we show that the effective potential has two minimum: one is outside $R_b$ and another one is inside the interior JMN-1 spacetime. One can check that this type of potential is not possible for JMN-2 spacetime for any value of $\lambda$. In fig.~(\ref{potbehav}), we show how the nature of effective potential changes when we change the $h$, where the effective potential corresponds to 
the spacetime structure which is internally JMN-1 and externally Schwarzschild spacetime. The fig.~(\ref{bb11}) show the situation when $  \sqrt{\frac{2-3M_0}{M_0}}>\frac{R_b}{h}$. In this case, there is no bound orbit inside the matching radius, however, in the exterior Schwarzschild spacetime particle's bound orbits exist. In fig.~(\ref{cc11}), we show the case where the inequality in eq.~(\ref{ineq1}) holds. The third case (shown in fig.~(\ref{aa11new})) is for $\frac{R_b}{h}>\frac{1}{\sqrt{3}M_0}$. In this case, bound orbit of the particle can only be possible inside the interior JMN-1 spacetime.

\section{Discussion and Conclusion}
In the present investigation, we examined the timelike particle trajectories in the JMN and Schwarzschild geometry, in order to explore the causal structure of these spacetimes, and to 
understand the main characteristic differences between the two. The study of these orbits bring out several interesting differences in the causal structure of these black hole and naked singularity spacetimes,
some of these are summarized as below: 
\begin{itemize}
    \item In the Schwarzschild spacetime, the bound trajectories of a massive particle always precess in the direction of the particle motion. However, in JMN naked singularity spacetimes, for a range of parameter values, the bound orbits precess in the opposite direction of particle motion, which is a very novel feature. In JMN-1 spacetime, for  $0<M_0<\frac13$, the orbits of a massive particle precess in the opposite direction of particle motion and for $\frac13<M_0<\frac23$, the orbits precess in the same direction of particle motion. In fig.~(\ref{precgen}), we show these two situations and compare it with the Schwarzschild spacetime. By the eq.~(\ref{m}), we show that there exists a maximum value of the angle ($\delta\phi_{prec}=254.56$ degree) for the opposite precession of a massive particle in JMN-1 spacetime.
    \item In Schwarzschild spacetime the perihelion point of a timelike orbit always lies in $r> 2M_0R_b$. On the other hand, in JMN naked singularity spacetime, the perihelion point of a timelike orbit lies in $r>0$ region.
    Therefore, in JMN spacetimes, a massive particle can go very close to the central naked singularity. In Schwarzschild spacetime, there exists an inner most stable circular orbit (ISCO) at $r=3M_0R_b$, however in JMN spacetimes, stable circular orbit of a massive particle can exist at any radius. This difference can create distinguishable accretion disk properties which can be detectable \cite{Joshi:2013dva},\cite{Shaikh:2018lcc}. 
    \item In fig.~(\ref{potbehav}), it can be seen that the effective potential of JMN spacetimes becomes positive infinite at the center. This indicates that in JMN spacetimes, a massive particle with non-zero angular momentum cannot reach the center. However, as the effective potential in Schwarzschild spacetime becomes negative infinity at the center, a massive particle with non-zero angular momentum and suitable total energy can reach the center.
\end{itemize}

As we know, GRAVITY and SINFONI are continuously observing the stellar motion around the Milkyway galaxy center. This observation can give us important information about the causal structure of galactic center, and on its mass and dynamics. In this context, we show in this paper that, the timelike geodesics in JMN naked singularity spacetime can be significantly different from the timelike geodesics in Schwarzschild spacetime. There are other astrophysical important naked singularity spacetimes (e.g. the JNW spacetime, Bertrand spacetime) for which we can do a similar comparison with Schwarzschild spacetimes, which will be reported separately \cite{parth}.

\end{document}